\documentclass[11pt]{article}
\usepackage{graphicx,amsmath,hyperref}

\newcommand{\mcm}[3]{\newcommand{#1}[#2]{{\ensuremath{#3}}}}

\usepackage{latexsym}
\usepackage{amssymb}

\mcm{\blank}{0}{(\emptybk)} \mcm{\dashbk}{0}{\mbox{---}}
\mcm{\emptybk}{0}{\:\:} \mcm{\hyph}{0}{\mbox{-}}

\mcm{\diagspace}{0}{\mbox{\hspace{2em}}}

\mcm{\cat}{1}{\mc{#1}} \mcm{\fcat}{1}{\mb{#1}}
\mcm{\mc}{1}{\mathcal{#1}} \mcm{\mr}{1}{\mathrm{#1}}
\mcm{\mi}{1}{\mathit{#1}} \mcm{\mb}{1}{\mathbf{#1}}
\mcm{\scat}{1}{\Bbb{#1}} \mcm{\twid}{1}{\widetilde{#1}}

\mcm{\elt}{0}{\in} \mcm{\sub}{0}{\,\subseteq\,}
\mcm{\such}{0}{\:|\:} \mcm{\without}{0}{\setminus}

\mcm{\atsr}{0}{\Box} \mcm{\eqv}{0}{\,\simeq\,}
\mcm{\iso}{0}{\,\cong\,}
\mcm{\of}{0}{\raisebox{0.2mm}{\ensuremath{\scriptstyle\circ}}}

\mcm{\bdry}{0}{\partial}

\mcm{\Bee}{0}{\cat{B}} \mcm{\Beep}{0}{\cat{B'}}
\mcm{\Eee}{0}{\cat{E}} \mcm{\Eeep}{0}{\cat{E'}}

\mcm{\Ess}{0}{\cat{S}} \mcm{\Tee}{0}{\cat{T}}
\mcm{\Teep}{0}{\cat{T'}} \mcm{\Stee}{0}{\scat{T}}
\mcm{\Steep}{0}{\scat{T'}}

\mcm{\blbk}{0}{\blank^{\blob}}
\mcm{\blob}{0}{\scriptscriptstyle{\bullet}}
\mcm{\stbk}{0}{\blank^{*}} \mcm{\ubl}{0}{{}^{\blob}}
\mcm{\ust}{0}{{}^{*}}

\mcm{\Cartpr}{0}{\pr{\Eee}{T}} \mcm{\Cartprp}{0}{\pr{\Eeep}{T'}}
\mcm{\Mnd}{0}{\triple{T}{\eta}{\mu}}
\mcm{\Zeropr}{0}{\pr{\Set}{\id}}

\mcm{\dopset}{0}{\ftrcat{\Delta^{\op}}{\Set}}
\mcm{\tropset}{0}{\ftrcat{\fcat{TR}^{\op}}{\Set}}

\mcm{\cod}{0}{\mr{cod}} \mcm{\dom}{0}{\mr{dom}}
\mcm{\End}{0}{\mr{End}} \mcm{\Hom}{0}{\mr{Hom}}
\mcm{\ob}{0}{\mr{ob}\,} \mcm{\op}{0}{\mr{op}}

\mcm{\comp}{0}{\mi{comp}} \mcm{\id}{0}{\mi{id}}
\mcm{\ids}{0}{\mi{ids}} \mcm{\mult}{0}{\mi{mult}}
\mcm{\unit}{0}{\mi{unit}}

\mcm{\Ab}{0}{\fcat{Ab}} \mcm{\Alg}{0}{\fcat{Alg}}
\mcm{\Bim}{1}{\fcat{Bim}(#1)} \mcm{\Cat}{0}{\fcat{Cat}}
\mcm{\Cay}{0}{\fcat{Cay}} \mcm{\Cpn}{1}{\pr{\Set/S_{#1}}{T_{#1}}}
\mcm{\fc}{0}{\fcat{fc}} \mcm{\fm}{0}{\fcat{fm}}
\mcm{\Graph}{0}{\fcat{Graph}} \mcm{\Gy}{0}{\fcat{Gy}}
\mcm{\Hpn}{1}{\pr{\Eee_{#1}}{P_{#1}}} \mcm{\Mon}{0}{\mb{Mon}}
\mcm{\Multicat}{0}{\fcat{Multicat}} \mcm{\One}{0}{\fcat{1}}
\mcm{\PD}{1}{\fcat{PD}_{#1}} \mcm{\Prof}{0}{\fcat{Prof}}
\mcm{\Set}{0}{\fcat{Set}} \mcm{\Span}{0}{\fcat{Span}}
\mcm{\Ssq}{0}{\fcat{Ssq}} \mcm{\Struc}{0}{\fcat{Struc}}
\mcm{\Sym}{0}{\fcat{Sym}} \mcm{\TR}{1}{\fcat{TR}(#1)}
\mcm{\Tr}{0}{\fcat{Tr}} \mcm{\Twocat}{0}{\fcat{2\hyph\Cat}}

\mcm{\integers}{0}{\mathbb{Z}}

\mcm{\range}{2}{#1,\,\ldots\,,#2}
\mcm{\bftuple}{2}{\tuplebts{\range{#1}{#2}}}
\mcm{\tuple}{3}{\tuplebts{\range{#1,#2}{#3}}}
\mcm{\rttuple}{1}{\tuplebts{\,\ldots\,,#1}}
\mcm{\abftuple}{2}{\atuplebts{\range{#1}{#2}}}
\mcm{\atuple}{3}{\atuplebts{\range{#1,#2}{#3}}}
\mcm{\arttuple}{1}{\atuplebts{\,\ldots\,,#1}}
\mcm{\sqbftuple}{2}{\obt\range{#1}{#2}\cbt}
\mcm{\pr}{2}{\tuplebts{#1,#2}}
\mcm{\triple}{3}{\tuplebts{#1,#2,#3}}

\mcm{\eend}{2}{#1[#2]} \mcm{\ehom}{3}{#1[#2,#3]}
\mcm{\ftrcat}{2}{[#1,#2]} \mcm{\homset}{3}{#1(#2,#3)}
\mcm{\multihom}{3}{#1(#2;#3)}
\mcm{\relhom}{5}{#1_{#2}(\range{#3}{#4};#5)}

\mcm{\go}{0}{\rTo} \mcm{\goby}{1}{\rTo^{#1}}
\mcm{\goesto}{0}{\,\longmapsto\,} \mcm{\goiso}{0}{\goby{\diso}}
\mcm{\monic}{0}{\rMonic} \mcm{\og}{0}{\lTo}
\mcm{\ogby}{1}{\lTo^{#1}}

\mcm{\gph}{2}{\spn{#1}{T #2}{#2}} \mcm{\graph}{4}{\spaan{#1}{T
#2}{#2}{#3}{#4}} \mcm{\oppair}{2}{\stackrel{\rTo^{#1}}{\lTo_{#2}}}
\mcm{\parpair}{2}{\stackrel{\rTo^{#1}}{\rTo_{#2}}}
\mcm{\spn}{3}{#2 \og #1 \go #3} \mcm{\spaan}{5}{#2 \ogby{#4} #1
\goby{#5} #3}

\mcm{\bktdvslob}{3}
    {\left(
    \begin{diagram}[height=1.5em]
    #1      \\
    \dTo>{\,#2} \\
    #3      \\
    \end{diagram}
    \right)}
\mcm{\slob}{3}{(#1 \goby{#2} #3)} \mcm{\vslob}{3}
    {\left.
    \begin{diagram}[height=1.5em]
    #1      \\
    \dTo>{\,#2} \\
    #3      \\
    \end{diagram}
    \right.}

\newenvironment{tree}
    {\begin{diagram}[height=1em,width=.75em,abut,noPS,tight]}
    {\end{diagram}}

\mcm{\enode}{0}{\circ}

\mcm{\nl}{1}{\stackrel{\textstyle #1}{\node}}
\mcm{\node}{0}{\bullet}

\mcm{\utree}{0}{\node}

\mcm{\diso}{0}{\sim}

\mcm{\vdiso}{0}{\wr}

\mcm{\nat}{0}{\mathbb{N}}

\mcm{\Onepr}{0}{\pr{\Graph}{\fc}}
\newlength{\nllwidth}
\newlength{\nllheight}
\newcommand{\stackbelow}[2]{%
\settowidth{\nllwidth}{\ensuremath{#1}\ensuremath{#2}}%
\settoheight{\nllheight}{\ensuremath{#2}}%
\addtolength{\nllheight}{.3ex}%
\mbox{%
\ensuremath{#1}%
\hspace{-.5\nllwidth}%
\raisebox{-1\nllheight}{\ensuremath{#2}}}}
\mcm{\nlal}{2}{\stackbelow{\nl{#1}}{#2}}
\mcm{\nll}{1}{\stackbelow{\node}{#1}} \mcm{\wun}{0}{\fcat{1}}
\mcm{\atuplebts}{1}{\langle #1 \rangle} \mcm{\tuplebts}{1}{(#1)}
\mcm{\bo}{0}{(} \mcm{\bc}{0}{)}
\mcm{\UBilax}{0}{\fcat{UBicat}_\mr{lax}}
\mcm{\UBiwk}{0}{\fcat{UBicat}_\mr{wk}}
\mcm{\UBistr}{0}{\fcat{UBicat}_\mr{str}}
\mcm{\Bilax}{0}{\fcat{Bicat}_\mr{lax}}
\mcm{\Biwk}{0}{\fcat{Bicat}_\mr{wk}}
\mcm{\Bistr}{0}{\fcat{Bicat}_\mr{str}} \mcm{\rotsub}{0}{\cup
\raisebox{0.1em}{$\scriptstyle{|}$}} \mcm{\pd}{0}{\fcat{pd}}
\mcm{\rep}{1}{\widehat{#1}} \mcm{\ovln}{1}{\overline{#1}}
\mcm{\Gph}{0}{\fcat{Gph}} \mcm{\tr}{0}{\fcat{tr}}

\mcm{\ladj}{0}{\,\dashv\,} \mcm{\zeropd}{0}{\node}
    {\end{diagram}}
\mcm{\END}{0}{\fcat{End}} \mcm{\HOM}{0}{\fcat{Hom}}

\newlength{\gwidth} 
\newlength{\gvert}  
\newlength{\gdrop}  
\newlength{\gbaredrop}  
\newlength{\goffset}    
\newlength{\gtemp}  

\newcommand{\present}[1]{%
\makebox[1\gwidth]{%
\rule[-1\gdrop]{0ex}{1\gvert}%
\raisebox{-1\gbaredrop}{#1}}}

\newcommand{\presentl}[1]{%
\makebox[1\gwidth][l]{%
\rule[-1\gdrop]{0ex}{1\gvert}%
\raisebox{-1\gbaredrop}{#1}}}

\newcommand{\presentr}[1]{%
\makebox[1\gwidth][r]{%
\rule[-1\gdrop]{0ex}{1\gvert}%
\raisebox{-1\gbaredrop}{#1}}}

\newcommand{\ginitdims}[2]{
\setlength{\unitlength}{1em}
\setlength{\goffset}{.25\unitlength}
\setlength{\gwidth}{#1\unitlength}
\setlength{\gvert}{#2\unitlength}
\setlength{\gdrop}{.5\gvert}
\addtolength{\gdrop}{-1\goffset}
\setlength{\gbaredrop}{1\gdrop}
\addtolength{\gvert}{.6\unitlength}
\addtolength{\gdrop}{.3\unitlength}}    

\newcommand{\cinitdims}[2]{
\setlength{\unitlength}{1em}
\setlength{\goffset}{.35\unitlength}
\setlength{\gwidth}{#1\unitlength}
\setlength{\gvert}{#2\unitlength}
\setlength{\gdrop}{.5\gvert}
\addtolength{\gdrop}{-1\goffset}
\setlength{\gbaredrop}{1\gdrop}
\addtolength{\gvert}{.6\unitlength}
\addtolength{\gdrop}{.3\unitlength}}    

\newcommand{\gsinitdims}[2]{
\setlength{\unitlength}{0.5em}
\setlength{\goffset}{.25\unitlength}
\setlength{\gwidth}{#1\unitlength}
\setlength{\gvert}{#2\unitlength}
\setlength{\gdrop}{.5\gvert}
\addtolength{\gdrop}{-1\goffset}
\setlength{\gbaredrop}{1\gdrop}
\addtolength{\gvert}{.6\unitlength}
\addtolength{\gdrop}{.3\unitlength}}    

\newcommand{\sidespic}[1]{%
\settowidth{\gtemp}{\ensuremath{#1}}%
\addtolength{\gwidth}{1\gtemp}}

\newcommand{\abovepic}[1]{%
\settoheight{\gtemp}{\ensuremath{#1}}%
\addtolength{\gvert}{1\gtemp}%
\settodepth{\gtemp}{\ensuremath{#1}}%
\addtolength{\gvert}{1\gtemp}}

\newcommand{\belowpic}[1]{%
\settoheight{\gtemp}{\ensuremath{#1}}%
\addtolength{\gvert}{1\gtemp}%
\addtolength{\gdrop}{1\gtemp}%
\settodepth{\gtemp}{\ensuremath{#1}}%
\addtolength{\gvert}{1\gtemp}%
\addtolength{\gdrop}{1\gtemp}}

\newcommand{\cell}[4]{\put(#1,#2){\makebox(0,0)[#3]{\ensuremath{#4}}}}
\mcm{\zmark}{0}{\scriptstyle{\bullet}}

\newcommand{\pregfst}[1]{%
\begin{picture}(0.5,0.2)(-0.5,-0.2)%
\cell{-0.1}{-0.2}{tr}{#1}%
\cell{0}{0}{c}{\zmark}%
\end{picture}}

\mcm{\gfst}{1}{%
\ginitdims{0.5}{0.4}%
\sidespic{#1}%
\belowpic{#1}%
\presentr{\pregfst{#1}}}

\newcommand{\preglst}[1]{%
\begin{picture}(0.5,0.2)(0,-0.2)%
\cell{0.1}{-0.2}{tl}{#1}%
\cell{0.05}{0}{c}{\zmark}%
\end{picture}}

\mcm{\glst}{1}{%
\ginitdims{.5}{.4}%
\sidespic{#1}%
\belowpic{#1}%
\presentl{\preglst{#1}}}

\newcommand{\preglft}[1]{%
\begin{picture}(0,0.2)(0,-0.2)%
\cell{-0.1}{-0.2}{tr}{#1}%
\cell{0.05}{0}{c}{\zmark}%
\end{picture}}

\mcm{\glft}{1}{%
\ginitdims{0}{.4}%
\belowpic{#1}%
\present{\preglft{#1}}}

\newcommand{\pregrgt}[1]{%
\begin{picture}(0,0.2)(0,-0.2)%
\cell{0.1}{-0.2}{tl}{#1}%
\cell{0.05}{0}{c}{\zmark}%
\end{picture}}

\mcm{\grgt}{1}{%
\ginitdims{0}{.4}%
\belowpic{#1}%
\present{\pregrgt{#1}}}

\newcommand{\pregblw}[1]{%
\begin{picture}(0,0.3)(0,-0.3)
\cell{0}{-0.3}{t}{#1}%
\cell{0.05}{0}{c}{\zmark}%
\end{picture}}

\mcm{\gblw}{1}{%
\ginitdims{0}{.6}%
\belowpic{#1}%
\present{\pregblw{#1}}}

\newcommand{\pregfbw}[1]{%
\begin{picture}(0,0.65)(0,-0.65)
\cell{0}{-0.65}{t}{#1}%
\cell{0.05}{0}{c}{\zmark}%
\end{picture}}

\mcm{\gfbw}{1}{%
\ginitdims{0}{1.3}%
\belowpic{#1}%
\present{\pregfbw{#1}}}

\newcommand{\pregzero}[1]{%
\begin{picture}(0.8,0.4)(-0.4,-0.4)
\cell{0}{-0.4}{t}{#1}%
\cell{0}{0}{c}{\zmark}%
\end{picture}}

\mcm{\gzero}{1}{%
\ginitdims{0.8}{.6}%
\belowpic{#1}%
\sidespic{#1}%
\present{\pregzero{#1}}}

\newcommand{\pregone}[1]{%
\begin{picture}(5,0.4)(0,-0.2)%
\cell{2.5}{0.2}{b}{#1}%
\put(0,0){\vector(1,0){5}}%
\end{picture}}

\mcm{\gone}{1}{%
\ginitdims{5}{0.4}%
\abovepic{#1}%
\present{\pregone{#1}}}

\newcommand{\pregtwo}[3]{%
\begin{picture}(5,3.4)(0,-0.2)%
\cell{2.5}{3.2}{b}{#1}%
\cell{2.5}{-.2}{t}{#2}%
\cell{2.7}{1.5}{l}{#3}%
\qbezier(0,1.5)(2.5,4.5)(5,1.5)%
\qbezier(0,1.5)(2.5,-1.5)(5,1.5)%
\put(5,1.5){\vector(1,-1){0}}%
\put(5,1.5){\vector(1,1){0}}%
\put(2.5,2.5){\vector(0,-1){2}}%
\end{picture}}

\mcm{\gtwo}{3}{%
\ginitdims{5}{3.4}%
\abovepic{#1}%
\belowpic{#2}%
\present{\pregtwo{#1}{#2}{#3}}}

\newcommand{\pregthree}[5]{%
\begin{picture}(5,5.4)(0,-1.2)%
\cell{2.5}{4.2}{b}{#1}%
\cell{1.5}{1.7}{b}{#2}%
\cell{2.5}{-1.2}{t}{#3}%
\cell{2.7}{2.75}{l}{#4}%
\cell{2.7}{0.25}{l}{#5}%
\qbezier(0,1.5)(2.5,6.5)(5,1.5)%
\qbezier(0,1.5)(2.5,-3.5)(5,1.5)%
\put(0,1.5){\vector(1,0){5}}%
\put(2.5,3.5){\vector(0,-1){1.5}}%
\put(2.5,1){\vector(0,-1){1.5}}%
\put(5,1.5){\vector(1,-3){0}}%
\put(5,1.5){\vector(1,3){0}}%
\end{picture}}

\mcm{\gthree}{5}{%
\ginitdims{5}{5.4}%
\abovepic{#1}%
\belowpic{#3}%
\present{\pregthree{#1}{#2}{#3}{#4}{#5}}}

\newcommand{\pregfour}[7]{%
\begin{picture}(5,8.4)(0,-2.7)%
\cell{2.5}{5.7}{b}{#1}%
\cell{1.5}{2.8}{b}{#2}%
\cell{1.5}{0.2}{t}{#3}%
\cell{2.5}{-2.7}{t}{#4}%
\cell{2.7}{4.25}{l}{#5}%
\cell{2.7}{1.5}{l}{#6}%
\cell{2.7}{-1.25}{l}{#7}%
\qbezier(0,1.5)(2.5,9.5)(5,1.5)%
\qbezier(0,1.5)(2.5,4)(5,1.5)%
\qbezier(0,1.5)(2.5,-1)(5,1.5)%
\qbezier(0,1.5)(2.5,-6.5)(5,1.5)%
\put(2.5,5.25){\vector(0,-1){2}}%
\put(2.5,2.5){\vector(0,-1){2}}%
\put(2.5,-0.25){\vector(0,-1){2}}%
\put(5,1.5){\vector(1,-4){0}}%
\put(5,1.5){\vector(4,-3){0}}%
\put(5,1.5){\vector(4,3){0}}%
\put(5,1.5){\vector(1,4){0}}%
\end{picture}}

\mcm{\gfour}{7}{%
\ginitdims{5}{8.4}%
\abovepic{#1}%
\belowpic{#4}%
\present{\pregfour{#1}{#2}{#3}{#4}{#5}{#6}{#7}}}

\newcommand{\pregthreecell}[5]{%
\begin{picture}(8,5)(-4,-2.5)%
\cell{0}{2.5}{b}{#1}%
\cell{0}{-2.5}{t}{#2}%
\cell{-1.7}{0}{r}{#3}%
\cell{1.7}{0}{l}{#4}%
\cell{0}{0.2}{b}{#5}%
\qbezier(-4,0)(0,4.2)(4,0)%
\qbezier(-4,0)(0,-4.2)(4,0)%
\qbezier(-0.5,1.8)(-2.5,0)(-0.5,-1.8)%
\qbezier(0.5,1.8)(2.5,0)(0.5,-1.8)%
\put(-1,0){\vector(1,0){2}}%
\put(4,0){\vector(1,-1){0}}%
\put(4,0){\vector(1,1){0}}%
\put(-0.5,-1.8){\vector(1,-1){0}}%
\put(0.5,-1.8){\vector(-1,-1){0}}%
\end{picture}}

\mcm{\gthreecell}{5}{%
\ginitdims{8}{5}%
\abovepic{#1}%
\belowpic{#2}%
\present{\pregthreecell{#1}{#2}{#3}{#4}{#5}}}

\newcommand{\pregthreecellu}{%
\begin{picture}(5,3.4)(-0.5,-0.2)%
\qbezier(-.5,1.5)(2,4.5)(4.5,1.5)%
\qbezier(-.5,1.5)(2,-1.5)(4.5,1.5)%
\qbezier(1.5,2.7)(0.5,1.5)(1.5,0.3)%
\qbezier(2.5,2.7)(3.5,1.5)(2.5,0.3)%
\put(1.3,1.5){\vector(1,0){1.4}}%
\put(4.5,1.5){\vector(1,-1){0}}%
\put(4.5,1.5){\vector(1,1){0}}%
\put(1.5,0.3){\vector(2,-3){0}}%
\put(2.5,0.3){\vector(-2,-3){0}}%
\end{picture}}

\mcm{\gthreecellu}{0}{%
\ginitdims{5}{3.4}%
\present{\pregthreecellu}}

\newcommand{\pregtwocentre}[3]{%
\begin{picture}(5,3.4)(0,-0.2)%
\cell{2.5}{3.2}{b}{#1}%
\cell{2.5}{-.2}{t}{#2}%
\cell{2.5}{1.5}{c}{#3}%
\qbezier(0,1.5)(2.5,4.5)(5,1.5)%
\qbezier(0,1.5)(2.5,-1.5)(5,1.5)%
\put(5,1.5){\vector(1,-1){0}}%
\put(5,1.5){\vector(1,1){0}}%
\put(2.5,2.5){\vector(0,-1){2}}%
\end{picture}}

\mcm{\gtwocentre}{3}{%
\ginitdims{5}{3.4}%
\abovepic{#1}%
\belowpic{#2}%
\present{\pregtwocentre{#1}{#2}{#3}}}

\newcommand{\pregspecialone}[9]{%
\begin{picture}(8,8)(-4,-4)%
\cell{0}{3.9}{b}{#1}%
\cell{-2}{-0.2}{t}{#2}%
\cell{0}{-3.9}{t}{#3}%
\cell{-1.5}{1.1}{r}{#4}%
\cell{0.2}{1.5}{l}{#5}%
\cell{1.5}{1.1}{l}{#6}%
\cell{0.2}{-2}{l}{#7}%
\cell{-0.9}{2.3}{b}{#8}%
\cell{0.9}{2.3}{b}{#9}%
\qbezier(-4,0)(0,8)(4,0)%
\qbezier(-4,0)(0,-8)(4,0)%
\qbezier(-0.5,3.4)(-3.5,2)(-0.5,0.6)%
\qbezier(0.5,3.4)(3.5,2)(0.5,0.6)%
\put(-4,0){\vector(1,0){8}}%
\put(0,3.4){\vector(0,-1){2.8}}%
\put(0,-0.8){\vector(0,-1){2.4}}%
\put(-1.5,2.2){\vector(1,0){1.2}}%
\put(0.3,2.2){\vector(1,0){1.2}}%
\put(4,0){\vector(1,-2){0}}%
\put(4,0){\vector(1,2){0}}%
\put(-0.5,0.6){\vector(2,-1){0}}%
\put(0.5,0.6){\vector(-2,-1){0}}%
\end{picture}}

\mcm{\gspecialone}{9}{%
\ginitdims{8}{8}%
\abovepic{#1}%
\belowpic{#3}%
\present{\pregspecialone{#1}{#2}{#3}{#4}{#5}{#6}{#7}{#8}{#9}}}

\newcommand{\pregspecialtwo}{%
\begin{picture}(5,3.4)(0,-0.2)%
\qbezier(0,1.5)(2.5,4.5)(5,1.5)%
\qbezier(0,1.5)(2.5,-1.5)(5,1.5)%
\qbezier(1.7,2.5)(0,1.5)(1.7,0.5)%
\qbezier(3.3,2.5)(5,1.5)(3.3,0.5)%
\put(5,1.5){\vector(1,-1){0}}%
\put(5,1.5){\vector(1,1){0}}%
\put(1.7,0.5){\vector(3,-2){0}}%
\put(3.3,0.5){\vector(-3,-2){0}}%
\put(2.5,2.5){\vector(0,-1){2}}%
\put(1.2,1.5){\vector(1,0){1}}%
\put(2.8,1.5){\vector(1,0){1}}%
\end{picture}}

\mcm{\gspecialtwo}{0}{%
\ginitdims{5}{3.4}%
\present{\pregspecialtwo}}

\newcommand{\pregspecialthree}{%
\begin{picture}(5,5.4)(0,-1.2)%
\qbezier(0,1.5)(2.5,6.5)(5,1.5)%
\qbezier(0,1.5)(2.5,-3.5)(5,1.5)%
\qbezier(2,3.5)(1,2.75)(2,2)%
\qbezier(3,3.5)(4,2.75)(3,2)%
\qbezier(2,1)(1,0.25)(2,-0.5)%
\qbezier(3,1)(4,0.25)(3,-0.5)%
\put(0,1.5){\vector(1,0){5}}%
\put(1.5,2.75){\vector(1,0){2}}%
\put(1.5,0.25){\vector(1,0){2}}%
\put(5,1.5){\vector(1,-3){0}}%
\put(5,1.5){\vector(1,3){0}}%
\put(2,2){\vector(1,-1){0}}%
\put(3,2){\vector(-1,-1){0}}%
\put(2,-0.5){\vector(1,-1){0}}%
\put(3,-0.5){\vector(-1,-1){0}}%
\end{picture}}

\mcm{\gspecialthree}{0}{%
\ginitdims{5}{5.4}%
\present{\pregspecialthree}}

\newcommand{\pregonew}[1]{%
\begin{picture}(8,0.4)(0,-0.2)%
\cell{4}{0.2}{b}{#1}%
\put(0,0){\vector(1,0){8}}%
\end{picture}}

\mcm{\gonew}{1}{%
\ginitdims{8}{0.4}%
\abovepic{#1}%
\present{\pregonew{#1}}}

\mcm{\gzersu}{0}{%
\gsinitdims{0}{.6}%
\present{\pregblw{}}}

\mcm{\gonesu}{0}{%
\gsinitdims{5}{0.4}%
\present{\pregone{}}}

\mcm{\gtwosu}{0}{%
\gsinitdims{5}{3.4}%
\present{\pregtwo{}{}{}}}

\mcm{\gthreesu}{0}{%
\gsinitdims{5}{5.4}%
\present{\pregthree{}{}{}{}{}}}

\mcm{\gfoursu}{0}{%
\gsinitdims{5}{8.4}%
\present{\pregfour{}{}{}{}{}{}{}}}

\newcommand{\precone}[1]{%
\begin{picture}(4.2,0.4)(-0.3,-0.2)%
\cell{1.8}{0.2}{b}{#1}%
\put(0,0){\vector(1,0){3.6}}%
\end{picture}}

\mcm{\cone}{1}{%
\cinitdims{4.2}{0.4}%
\abovepic{#1}%
\present{\precone{#1}}}

\mcm{\gfstsu}{0}{%
\gsinitdims{0.5}{0.4}%
\presentr{\pregfst{}}}

\mcm{\glstsu}{0}{%
\gsinitdims{0.5}{0.4}%
\presentl{\preglst{}}}

\newcommand{\prectwodbl}[3]%
{\begin{picture}(4.2,3.4)(-0.1,-0.2)%
\cell{2}{3.2}{b}{#1}%
\cell{2}{-0.2}{t}{#2}%
\cell{2.3}{1.5}{l}{#3}%
\qbezier(0,2)(2,4)(4,2)%
\qbezier(0,1)(2,-1)(4,1)%
\put(4,2){\vector(1,-1){0}}%
\put(4,1){\vector(1,1){0}}%
\put(1.9,2.5){\line(0,-1){1.8}}%
\put(2.1,2.5){\line(0,-1){1.8}}%
\cell{2.01}{0.4}{b}{\vee}%
\end{picture}}

\mcm{\ctwodbl}{3}{%
\cinitdims{4.2}{3.4}%
\abovepic{#1}%
\belowpic{#2}%
\present{\prectwodbl{#1}{#2}{#3}}}

\newcommand{\precthreedbl}[5]{%
\begin{picture}(4.2,5.4)(-0.1,-0.2)%
\cell{2}{5.2}{b}{#1}%
\cell{1}{2.7}{b}{#2}%
\cell{2}{-.2}{t}{#3}%
\cell{2.3}{3.75}{l}{#4}%
\cell{2.3}{1.25}{l}{#5}%
\qbezier(0,3)(2,7)(4,3)%
\qbezier(0,2)(2,-2)(4,2)%
\put(0,2.5){\vector(1,0){4}}%
\put(1.9,4.5){\line(0,-1){1.3}}%
\put(2.1,4.5){\line(0,-1){1.3}}%
\cell{2.01}{2.9}{b}{\vee}%
\put(1.9,2){\line(0,-1){1.3}}%
\put(2.1,2){\line(0,-1){1.3}}%
\cell{2.01}{0.4}{b}{\vee}%
\put(4,3){\vector(1,-3){0}}%
\put(4,2){\vector(1,3){0}}%
\end{picture}}

\mcm{\cthreedbl}{5}{%
\cinitdims{4.2}{5.4}%
\abovepic{#1}%
\belowpic{#3}%
\present{\precthreedbl{#1}{#2}{#3}{#4}{#5}}}

\newcommand{\precthreecelltrp}[5]{%
\begin{picture}(8.2,5)(-4.1,-2.5)%
\cell{0}{2.5}{b}{#1}%
\cell{0}{-2.5}{t}{#2}%
\cell{-1.8}{0}{r}{#3}%
\cell{1.8}{0}{l}{#4}%
\cell{0}{0.3}{b}{#5}%
\qbezier(-4,0.5)(0,4)(4,0.5)%
\qbezier(-4,-0.5)(0,-4)(4,-0.5)%
\qbezier(-0.6,2)(-2.6,0)(-0.6,-2)%
\qbezier(-0.4,2)(-2.4,0)(-0.5,-1.9)%
\cell{-0.6}{-2}{b}{\lrcorner}%
\qbezier(0.4,2)(2.4,0)(0.5,-1.9)%
\qbezier(0.6,2)(2.6,0)(0.6,-2)%
\cell{0.65}{-2}{b}{\llcorner}%
\put(-1,0.15){\line(1,0){1.7}}%
\put(-1,0){\line(1,0){2}}%
\put(-1,-0.15){\line(1,0){1.7}}%
\cell{1.15}{0}{r}{>}%
\put(4,0.5){\vector(1,-1){0}}%
\put(4,-0.5){\vector(1,1){0}}%
\end{picture}}

\mcm{\cthreecelltrp}{5}{%
\cinitdims{8.2}{5}%
\abovepic{#1}%
\belowpic{#2}%
\present{\precthreecelltrp{#1}{#2}{#3}{#4}{#5}}}

\newcommand{\prectwo}[3]%
{\begin{picture}(4.2,3.4)(-0.1,-0.2)%
\cell{2}{3.2}{b}{#1}%
\cell{2}{-0.2}{t}{#2}%
\cell{2.2}{1.5}{l}{#3}%
\qbezier(0,2)(2,4)(4,2)%
\qbezier(0,1)(2,-1)(4,1)%
\put(4,2){\vector(1,-1){0}}%
\put(4,1){\vector(1,1){0}}%
\put(2,2.5){\vector(0,-1){2}}%
\end{picture}}

\mcm{\ctwo}{3}{%
\cinitdims{4.2}{3.4}%
\abovepic{#1}%
\belowpic{#2}%
\present{\prectwo{#1}{#2}{#3}}}

\newcommand{\precthree}[5]{%
\begin{picture}(4.2,5.4)(-0.1,-0.2)%
\cell{2}{5.2}{b}{#1}%
\cell{1}{2.7}{b}{#2}%
\cell{2}{-.2}{t}{#3}%
\cell{2.2}{3.75}{l}{#4}%
\cell{2.2}{1.25}{l}{#5}%
\qbezier(0,3)(2,7)(4,3)%
\qbezier(0,2)(2,-2)(4,2)%
\put(0,2.5){\vector(1,0){4}}%
\put(2,4.5){\vector(0,-1){1.5}}%
\put(2,2){\vector(0,-1){1.5}}%
\put(4,3){\vector(1,-3){0}}%
\put(4,2){\vector(1,3){0}}%
\end{picture}}

\mcm{\cthree}{5}{%
\cinitdims{4.2}{5.4}%
\abovepic{#1}%
\belowpic{#3}%
\present{\precthree{#1}{#2}{#3}{#4}{#5}}}

\newcommand{\prectwoop}[3]%
{\begin{picture}(4.2,3.4)(-0.1,-0.2)%
\cell{2}{3.2}{b}{#1}%
\cell{2}{-0.2}{t}{#2}%
\cell{2.2}{1.5}{l}{#3}%
\qbezier(0,2)(2,4)(4,2)%
\qbezier(0,1)(2,-1)(4,1)%
\put(0,2){\vector(-1,-1){0}}%
\put(0,1){\vector(-1,1){0}}%
\put(2,2.5){\vector(0,-1){2}}%
\end{picture}}

\mcm{\ctwoop}{3}{%
\cinitdims{4.2}{3.4}%
\abovepic{#1}%
\belowpic{#2}%
\present{\prectwoop{#1}{#2}{#3}}}

\newcommand{\prectwopar}[4]{%
\begin{picture}(4.2,3.4)(-0.1,-0.2)%
\cell{2}{3.2}{b}{#1}%
\cell{2}{-0.2}{t}{#2}%
\cell{1.6}{1.5}{r}{#3}%
\cell{2.4}{1.5}{l}{#4}%
\qbezier(0,2)(2,4)(4,2)%
\qbezier(0,1)(2,-1)(4,1)%
\put(4,2){\vector(1,-1){0}}%
\put(4,1){\vector(1,1){0}}%
\put(1.8,2.5){\vector(0,-1){2}}%
\put(2.2,2.5){\vector(0,-1){2}}%
\end{picture}}

\mcm{\ctwopar}{4}{%
\cinitdims{4.2}{3.4}%
\abovepic{#1}%
\belowpic{#2}%
\present{\prectwopar{#1}{#2}{#3}{#4}}}

\newcommand{\precthreein}[5]{%
\begin{picture}(4.2,5.4)(-0.1,-0.2)%
\cell{2}{5.2}{b}{#1}%
\cell{1}{2.7}{b}{#2}%
\cell{2}{-.2}{t}{#3}%
\cell{2.2}{3.75}{l}{#4}%
\cell{2.2}{1.25}{l}{#5}%
\qbezier(0,3)(2,7)(4,3)%
\qbezier(0,2)(2,-2)(4,2)%
\put(0,2.5){\vector(1,0){4}}%
\put(2,4.5){\vector(0,-1){1.5}}%
\put(2,0.5){\vector(0,1){1.5}}%
\put(4,3){\vector(1,-3){0}}%
\put(4,2){\vector(1,3){0}}%
\end{picture}}

\mcm{\cthreein}{5}{%
\cinitdims{4.2}{5.4}%
\abovepic{#1}%
\belowpic{#3}%
\present{\precthreein{#1}{#2}{#3}{#4}{#5}}}

\newcommand{\precthreecell}[5]{%
\begin{picture}(8.2,5)(-4.1,-2.5)%
\cell{0}{2.5}{b}{#1}%
\cell{0}{-2.5}{t}{#2}%
\cell{-1.7}{0}{r}{#3}%
\cell{1.7}{0}{l}{#4}%
\cell{0}{0.2}{b}{#5}%
\qbezier(-4,0.5)(0,4)(4,0.5)%
\qbezier(-4,-0.5)(0,-4)(4,-0.5)%
\qbezier(-0.5,2)(-2.5,0)(-0.5,-2)%
\qbezier(0.5,2)(2.5,0)(0.5,-2)%
\put(-1,0){\vector(1,0){2}}%
\put(4,0.5){\vector(1,-1){0}}%
\put(4,-0.5){\vector(1,1){0}}%
\put(-0.5,-2){\vector(1,-1){0}}%
\put(0.5,-2){\vector(-1,-1){0}}%
\end{picture}}

\mcm{\cthreecell}{5}{%
\cinitdims{8.2}{5}%
\abovepic{#1}%
\belowpic{#2}%
\present{\precthreecell{#1}{#2}{#3}{#4}{#5}}}

\newcommand{\precthreecellpar}[6]{%
\begin{picture}(8.2,5)(-4.1,-2.5)%
\cell{0}{2.5}{b}{#1}%
\cell{0}{-2.5}{t}{#2}%
\cell{-1.7}{0}{r}{#3}%
\cell{1.7}{0}{l}{#4}%
\cell{0}{0.4}{b}{#5}%
\cell{0}{-0.4}{t}{#6}%
\qbezier(-4,0.5)(0,4)(4,0.5)%
\qbezier(-4,-0.5)(0,-4)(4,-0.5)%
\qbezier(-0.5,2)(-2.5,0)(-0.5,-2)%
\qbezier(0.5,2)(2.5,0)(0.5,-2)%
\put(-1,0.2){\vector(1,0){2}}%
\put(-1,-0.2){\vector(1,0){2}}%
\put(4,0.5){\vector(1,-1){0}}%
\put(4,-0.5){\vector(1,1){0}}%
\put(-0.5,-2){\vector(1,-1){0}}%
\put(0.5,-2){\vector(-1,-1){0}}%
\end{picture}}

\mcm{\cthreecellpar}{6}{%
\cinitdims{8.2}{5}%
\abovepic{#1}%
\belowpic{#2}%
\present{\precthreecellpar{#1}{#2}{#3}{#4}{#5}{#6}}}

\newcommand{\prectwov}[5]{%
\begin{picture}(3.4,4.2)(0.8,0.9)%
\cell{2.5}{5.1}{b}{#1}%
\cell{2.5}{0.9}{t}{#2}%
\cell{0.8}{3}{r}{#3}%
\cell{4.2}{3}{l}{#4}%
\cell{2.5}{3.2}{b}{#5}%
\qbezier(2,5)(0,3)(2,1)%
\qbezier(3,5)(5,3)(3,1)%
\put(2,1){\vector(1,-1){0}}%
\put(3,1){\vector(-1,-1){0}}%
\put(1.5,3){\vector(1,0){2}}%
\end{picture}}

\mcm{\ctwov}{5}{%
\cinitdims{3.4}{4.2}%
\abovepic{#1}%
\belowpic{#2}%
\sidespic{#3}%
\sidespic{#4}%
\present{\prectwov{#1}{#2}{#3}{#4}{#5}}}

\newcommand{\precthreecellv}[7]{%
\begin{picture}(5,8.2)(0.5,-1.6)%
\cell{3}{6.6}{b}{#1}%
\cell{3}{-1.6}{t}{#2}%
\cell{0.5}{2.5}{r}{#3}%
\cell{5.5}{2.5}{l}{#4}%
\cell{3}{4.2}{b}{#5}%
\cell{3}{0.8}{t}{#6}%
\cell{3.2}{2.5}{l}{#7}%
\qbezier(3.5,6.5)(7,2.5)(3.5,-1.5)%
\qbezier(2.5,6.5)(-1,2.5)(2.5,-1.5)%
\put(2.5,-1.5){\vector(1,-1){0}}%
\put(3.5,-1.5){\vector(-1,-1){0}}%
\qbezier(1,3)(3,5)(5,3)%
\qbezier(1,2)(3,0)(5,2)%
\put(5,3){\vector(1,-1){0}}%
\put(5,2){\vector(1,1){0}}%
\put(3,3.5){\vector(0,-1){2}}%
\end{picture}}

\mcm{\cthreecellv}{7}{%
\cinitdims{5}{8.2}%
\abovepic{#1}%
\belowpic{#2}%
\sidespic{#3}%
\sidespic{#4}%
\present{\precthreecellv{#1}{#2}{#3}{#4}{#5}{#6}{#7}}}

\newcommand{\pretopez}[2]{%
\begin{picture}(2.6,2.3)(-1.3,-2.2)%
\cell{0}{-2.2}{t}{#1}%
\cell{0}{-1.2}{c}{#2}%
\qbezier(0,0)(-2,-2)(0,-2)%
\qbezier(0,0)(2,-2)(0,-2)%
\put(0,0){\vector(-1,1){0}}%
\end{picture}}

\mcm{\topez}{2}{%
\ginitdims{2.6}{2.3}%
\belowpic{#1}%
\present{\pretopez{#1}{#2}}}

\newcommand{\pretopea}[3]{%
\begin{picture}(4,1.9)(-2,-0,2)%
\cell{0}{1.7}{b}{#1}%
\cell{0}{-0.2}{t}{#2}%
\cell{0}{0.7}{c}{#3}%
\qbezier(-2,0)(0,3)(2,0)%
\put(-2,0){\vector(1,0){4}}%
\put(2,0){\vector(2,-3){0}}%
\end{picture}}

\mcm{\topea}{3}{%
\ginitdims{4}{1.9}%
\abovepic{#1}%
\belowpic{#2}%
\present{\pretopea{#1}{#2}{#3}}}

\newcommand{\pretopeb}[4]{%
\begin{picture}(4,2.2)(-2,-0.2)%
\cell{-1.1}{1}{br}{#1}%
\cell{1.1}{1}{bl}{#2}%
\cell{0}{-0.2}{t}{#3}%
\cell{0}{0.8}{c}{#4}%
\put(-2,0){\vector(1,1){2}}%
\put(0,2){\vector(1,-1){2}}%
\put(-2,0){\vector(1,0){4}}%
\end{picture}}

\mcm{\topeb}{4}{%
\ginitdims{4}{2.2}%
\belowpic{#3}%
\present{\pretopeb{#1}{#2}{#3}{#4}}}

\newcommand{\pretopec}[5]{%
\begin{picture}(4,2.2)(-2,-0.2)%
\cell{-1.8}{1}{br}{#1}%
\cell{0}{2.2}{b}{#2}%
\cell{1.8}{1}{bl}{#3}%
\cell{0}{-0.2}{t}{#4}%
\cell{0}{0.8}{c}{#5}%
\put(-2,0){\vector(1,2){1}}%
\put(-1,2){\vector(1,0){2}}%
\put(1,2){\vector(1,-2){1}}%
\put(-2,0){\vector(1,0){4}}%
\end{picture}}

\mcm{\topec}{5}{%
\ginitdims{4}{2.2}%
\sidespic{#1}%
\abovepic{#2}%
\sidespic{#3}%
\belowpic{#4}%
\present{\pretopec{#1}{#2}{#3}{#4}{#5}}}

\newcommand{\pretoped}[6]{%
\begin{picture}(4,2.5)(-2,-0.2)%
\cell{-2}{0.6}{br}{#1}%
\cell{-0.7}{2.2}{br}{#2}%
\cell{0.7}{2.2}{bl}{#3}%
\cell{2}{0.6}{bl}{#4}%
\cell{0}{-0.2}{t}{#5}%
\cell{0}{0.8}{c}{#6}%
\put(-2,0){\vector(1,3){0.5}}%
\put(-1.5,1.5){\vector(3,2){1.5}}%
\put(0,2.5){\vector(3,-2){1.5}}%
\put(1.5,1.5){\vector(1,-3){0.5}}%
\put(-2,0){\vector(1,0){4}}%
\end{picture}}

\mcm{\toped}{6}{%
\ginitdims{4}{2.5}%
\sidespic{#1}%
\abovepic{#2}%
\abovepic{#3}%
\sidespic{#4}%
\belowpic{#5}%
\present{\pretoped{#1}{#2}{#3}{#4}{#5}{#6}}}

\newcommand{\pretopeq}[5]{%
\begin{picture}(4,2.5)(-2,-0.2)%
\cell{-2}{0.6}{br}{#1}%
\cell{-1}{2.2}{br}{#2}%
\cell{2}{0.6}{bl}{#3}%
\cell{0}{-0.2}{t}{#4}%
\cell{0}{0.8}{c}{#5}%
\put(-2,0){\vector(1,3){0.5}}%
\put(-1.5,1.5){\vector(1,1){1}}%
\cell{0.9}{2.3}{c}{\ddots}
\put(1.5,1.5){\vector(1,-3){0.5}}%
\put(-2,0){\vector(1,0){4}}%
\end{picture}}

\mcm{\topeq}{5}{%
\ginitdims{4}{2.5}%
\sidespic{#1}%
\abovepic{#2}%
\sidespic{#3}%
\belowpic{#4}%
\present{\pretopeq{#1}{#2}{#3}{#4}{#5}}}

\newcommand{\pretopebase}[1]{%
\begin{picture}(4,0.4)(0,-0.2)%
\cell{2}{0.2}{b}{#1}%
\put(0,0){\vector(1,0){4}}%
\end{picture}}

\mcm{\topebase}{1}{%
\ginitdims{4}{0.4}%
\abovepic{#1}%
\present{\pretopebase{#1}}}

\newcommand{\pretopezs}[2]{%
\begin{picture}(2.6,2.3)(-1.3,-2.2)%
\cell{0}{-2.2}{t}{#1}%
\cell{0}{-1.2}{c}{#2}%
\qbezier(0,0)(-2,-2)(0,-2)%
\qbezier(0,0)(2,-2)(0,-2)%
\end{picture}}

\mcm{\topezs}{2}{%
\ginitdims{2.6}{2.3}%
\belowpic{#1}%
\present{\pretopezs{#1}{#2}}}

\newcommand{\pretopeas}[3]{%
\begin{picture}(4,1.9)(-2,-0,2)%
\cell{0}{1.7}{b}{#1}%
\cell{0}{-0.2}{t}{#2}%
\cell{0}{0.7}{c}{#3}%
\qbezier(-2,0)(0,3)(2,0)%
\put(-2,0){\line(1,0){4}}%
\end{picture}}

\mcm{\topeas}{3}{%
\ginitdims{4}{1.9}%
\abovepic{#1}%
\belowpic{#2}%
\present{\pretopeas{#1}{#2}{#3}}}

\newcommand{\pretopebs}[4]{%
\begin{picture}(4,2.2)(-2,-0.2)%
\cell{-1.1}{1}{br}{#1}%
\cell{1.1}{1}{bl}{#2}%
\cell{0}{-0.2}{t}{#3}%
\cell{0}{0.8}{c}{#4}%
\put(-2,0){\line(1,1){2}}%
\put(0,2){\line(1,-1){2}}%
\put(-2,0){\line(1,0){4}}%
\end{picture}}

\mcm{\topebs}{4}{%
\ginitdims{4}{2.2}%
\belowpic{#3}%
\present{\pretopebs{#1}{#2}{#3}{#4}}}

\newcommand{\pretopecs}[5]{%
\begin{picture}(4,2.2)(-2,-0.2)%
\cell{-1.8}{1}{br}{#1}%
\cell{0}{2.2}{b}{#2}%
\cell{1.8}{1}{bl}{#3}%
\cell{0}{-0.2}{t}{#4}%
\cell{0}{0.8}{c}{#5}%
\put(-2,0){\line(1,2){1}}%
\put(-1,2){\line(1,0){2}}%
\put(1,2){\line(1,-2){1}}%
\put(-2,0){\line(1,0){4}}%
\end{picture}}

\mcm{\topecs}{5}{%
\ginitdims{4}{2.2}%
\sidespic{#1}%
\abovepic{#2}%
\sidespic{#3}%
\belowpic{#4}%
\present{\pretopecs{#1}{#2}{#3}{#4}{#5}}}

\newcommand{\pretopeds}[6]{%
\begin{picture}(4,2.5)(-2,-0.2)%
\cell{-2}{0.6}{br}{#1}%
\cell{-0.7}{2.2}{br}{#2}%
\cell{0.7}{2.2}{bl}{#3}%
\cell{2}{0.6}{bl}{#4}%
\cell{0}{-0.2}{t}{#5}%
\cell{0}{0.8}{c}{#6}%
\put(-2,0){\line(1,3){0.5}}%
\put(-1.5,1.5){\line(3,2){1.5}}%
\put(0,2.5){\line(3,-2){1.5}}%
\put(1.5,1.5){\line(1,-3){0.5}}%
\put(-2,0){\line(1,0){4}}%
\end{picture}}

\mcm{\topeds}{6}{%
\ginitdims{4}{2.5}%
\sidespic{#1}%
\abovepic{#2}%
\abovepic{#3}%
\sidespic{#4}%
\belowpic{#5}%
\present{\pretopeds{#1}{#2}{#3}{#4}{#5}{#6}}}

\newcommand{\pretopeqs}[5]{%
\begin{picture}(4,2.5)(-2,-0.2)%
\cell{-2}{0.6}{br}{#1}%
\cell{-1}{2.2}{br}{#2}%
\cell{2}{0.6}{bl}{#3}%
\cell{0}{-0.2}{t}{#4}%
\cell{0}{0.8}{c}{#5}%
\put(-2,0){\line(1,3){0.5}}%
\put(-1.5,1.5){\line(1,1){1}}%
\cell{0.9}{2.3}{c}{\ddots}
\put(1.5,1.5){\line(1,-3){0.5}}%
\put(-2,0){\line(1,0){4}}%
\end{picture}}

\mcm{\topeqs}{5}{%
\ginitdims{4}{2.5}%
\sidespic{#1}%
\abovepic{#2}%
\sidespic{#3}%
\belowpic{#4}%
\present{\pretopeqs{#1}{#2}{#3}{#4}{#5}}}

\newcommand{\pretopebases}[1]{%
\begin{picture}(4,0.4)(0,-0.2)%
\cell{2}{0.2}{b}{#1}%
\put(0,0){\line(1,0){4}}%
\end{picture}}

\mcm{\topebases}{1}{%
\ginitdims{4}{0.4}%
\abovepic{#1}%
\present{\pretopebases{#1}}}

\newcommand{\pregdots}[6]{%
\begin{picture}(5,8.4)(0,-2.7)%
\cell{2.5}{5.7}{b}{#1}%
\cell{1.5}{2.8}{b}{#2}%
\cell{1.5}{0.2}{t}{#3}%
\cell{2.5}{-2.7}{t}{#4}%
\cell{2.7}{4.25}{l}{#5}%
\cell{2.7}{-1.25}{l}{#6}%
\qbezier(0,1.5)(2.5,9.5)(5,1.5)%
\qbezier(0,1.5)(2.5,4)(5,1.5)%
\qbezier(0,1.5)(2.5,-1)(5,1.5)%
\qbezier(0,1.5)(2.5,-6.5)(5,1.5)%
\put(2.5,5.25){\vector(0,-1){2}}%
\put(2.5,-0.25){\vector(0,-1){2}}%
\cell{2.5}{1.7}{c}{\vdots}%
\put(5,1.5){\vector(1,-4){0}}%
\put(5,1.5){\vector(4,-3){0}}%
\put(5,1.5){\vector(4,3){0}}%
\put(5,1.5){\vector(1,4){0}}%
\end{picture}}

\mcm{\gdots}{6}{%
\ginitdims{5}{8.4}%
\abovepic{#1}%
\belowpic{#4}%
\present{\pregdots{#1}{#2}{#3}{#4}{#5}{#6}}}

\newlength{\volt}
\makeatletter

\def\diagram{\m@th\leftwidth=\z@ \rightwidth=\z@ \topheight=\z@
\botheight=\z@ \setbox\@picbox\hbox\bgroup}

\def\enddiagram{\egroup\wd\@picbox\rightwidth\unitlength
\ht\@picbox\topheight\unitlength \dp\@picbox\botheight\unitlength
\hskip\leftwidth\unitlength\box\@picbox}

\def\bfig{\begin{diagram}}
\def\efig{\end{diagram}}
\newcount\wideness \newcount\leftwidth \newcount\rightwidth
\newcount\highness \newcount\topheight \newcount\botheight

\def\ratchet#1#2{\ifnum#1<#2 \global #1=#2 \fi}

\def\putbox(#1,#2)#3{%
\horsize{\wideness}{#3} \divide\wideness by 2 {\advance\wideness
by #1 \ratchet{\rightwidth}{\wideness}} {\advance\wideness by -#1
\ratchet{\leftwidth}{\wideness}} \vertsize{\highness}{#3}
\divide\highness by 2 {\advance\highness by #2
\ratchet{\topheight}{\highness}} {\advance\highness by -#2
\ratchet{\botheight}{\highness}} \put(#1,#2){\makebox(0,0){$#3$}}}

\def\putlbox(#1,#2)#3{%
\horsize{\wideness}{#3} {\advance\wideness by #1
\ratchet{\rightwidth}{\wideness}} {\ratchet{\leftwidth}{-#1}}
\vertsize{\highness}{#3} \divide\highness by 2 {\advance\highness
by #2 \ratchet{\topheight}{\highness}} {\advance\highness by -#2
\ratchet{\botheight}{\highness}}
\put(#1,#2){\makebox(0,0)[l]{$#3$}}}

\def\putrbox(#1,#2)#3{%
\horsize{\wideness}{#3} {\ratchet{\rightwidth}{#1}}
{\advance\wideness by -#1 \ratchet{\leftwidth}{\wideness}}
\vertsize{\highness}{#3} \divide\highness by 2 {\advance\highness
by #2 \ratchet{\topheight}{\highness}} {\advance\highness by -#2
\ratchet{\botheight}{\highness}}
\put(#1,#2){\makebox(0,0)[r]{$#3$}}}

\def\adjust[#1]{} 

\newcount \coefa
\newcount \coefb
\newcount \coefc
\newcount\tempcounta
\newcount\tempcountb
\newcount\tempcountc
\newcount\tempcountd
\newcount\xext
\newcount\yext
\newcount\xoff
\newcount\yoff
\newcount\gap%
\newcount\arrowtypea
\newcount\arrowtypeb
\newcount\arrowtypec
\newcount\arrowtyped
\newcount\arrowtypee
\newcount\height
\newcount\width
\newcount\xpos
\newcount\ypos
\newcount\run
\newcount\rise
\newcount\arrowlength
\newcount\halflength
\newcount\arrowtype
\newdimen\tempdimen
\newdimen\xlen
\newdimen\ylen
\newsavebox{\tempboxa}%
\newsavebox{\tempboxb}%
\newsavebox{\tempboxc}%

\newdimen\w@dth

\def\setw@dth#1#2{\setbox\z@\hbox{\m@th$#1$}\w@dth=\wd\z@
\setbox\@ne\hbox{\m@th$#2$}\ifnum\w@dth<\wd\@ne \w@dth=\wd\@ne \fi
\advance\w@dth by 1.2em}

\def\t@^#1_#2{\allowbreak\def\n@one{#1}\def\n@two{#2}\mathrel
{\setw@dth{#1}{#2} \mathop{\hbox to
\w@dth{\rightarrowfill}}\limits \ifx\n@one\empty\else
^{\box\z@}\fi \ifx\n@two\empty\else _{\box\@ne}\fi}}
\def\t@@^#1{\@ifnextchar_{\t@^{#1}}{\t@^{#1}_{}}}
\def\to{\@ifnextchar^{\t@@}{\t@@^{}}}

\def\t@left^#1_#2{\def\n@one{#1}\def\n@two{#2}\mathrel{\setw@dth{#1}{#2}
\mathop{\hbox to \w@dth{\leftarrowfill}}\limits
\ifx\n@one\empty\else ^{\box\z@}\fi \ifx\n@two\empty\else
_{\box\@ne}\fi}}
\def\t@@left^#1{\@ifnextchar_{\t@left^{#1}}{\t@left^{#1}_{}}}
\def\toleft{\@ifnextchar^{\t@@left}{\t@@left^{}}}

\def\two@^#1_#2{\allowbreak
\def\n@one{#1}\def\n@two{#2}\mathrel{\setw@dth{#1}{#2}
\mathop{\vcenter{\lineskip\z@\baselineskip\z@
                 \hbox to \w@dth{\rightarrowfill}%
                 \hbox to \w@dth{\rightarrowfill}}%
       }\limits
\ifx\n@one\empty\else ^{\box\z@}\fi \ifx\n@two\empty\else
_{\box\@ne}\fi}}
\def\tw@@^#1{\@ifnextchar _{\two@^{#1}}{\two@^{#1}_{}}}
\def\two{\@ifnextchar ^{\tw@@}{\tw@@^{}}}

\def\tofr@^#1_#2{\def\n@one{#1}\def\n@two{#2}\mathrel{\setw@dth{#1}{#2}
\mathop{\vcenter{\hbox to \w@dth{\rightarrowfill}\kern-1.7ex
                 \hbox to \w@dth{\leftarrowfill}}%
       }\limits
\ifx\n@one\empty\else ^{\box\z@}\fi \ifx\n@two\empty\else
_{\box\@ne}\fi}}
\def\t@fr@^#1{\@ifnextchar_ {\tofr@^{#1}}{\tofr@^{#1}_{}}}
\def\tofro{\@ifnextchar^ {\t@fr@}{\t@fr@^{}}}

\def\mon{\mathop{\m@th\hbox to
      14.6\P@{\lasyb\char'51\hskip-2.1\P@$\arrext$\hss
$\mathord\rightarrow$}}\limits} 
\def\leftmono{\mathrel{\m@th\hbox to
14.6\P@{$\mathord\leftarrow$\hss$\arrext$\hskip-2.1\P@\lasyb\char'50%
}}\limits} 
\mathchardef\arrext="0200       

\def\settypes(#1,#2,#3){\arrowtypea#1 \arrowtypeb#2 \arrowtypec#3}
\def\settoheight#1#2{\setbox\@tempboxa\hbox{#2}#1\ht\@tempboxa\relax}%
\def\settodepth#1#2{\setbox\@tempboxa\hbox{#2}#1\dp\@tempboxa\relax}%
\def\settokens`#1`#2`#3`#4`{%
     \def\tokena{#1}\def\tokenb{#2}\def\tokenc{#3}\def\tokend{#4}}
\def\setsqparms[#1`#2`#3`#4;#5`#6]{%
\arrowtypea #1 \arrowtypeb #2 \arrowtypec #3 \arrowtyped #4
\width #5 \height #6 }
\def\setpos(#1,#2){\xpos=#1 \ypos#2}

\def\settriparms[#1`#2`#3;#4]{\settripairparms[#1`#2`#3`1`1;#4]}%

\def\settripairparms[#1`#2`#3`#4`#5;#6]{%
\arrowtypea #1 \arrowtypeb #2 \arrowtypec #3 \arrowtyped #4
\arrowtypee #5 \width #6 \height #6 }

\def\resetparms{\settripairparms[1`1`1`1`1;500]\width 500}

\resetparms

\def\mvector(#1,#2)#3{
\put(0,0){\vector(#1,#2){#3}}%
\put(0,0){\vector(#1,#2){26}}%
}
\def\evector(#1,#2)#3{{
\arrowlength #3
\put(0,0){\vector(#1,#2){\arrowlength}}%
\advance \arrowlength by-30
\put(0,0){\vector(#1,#2){\arrowlength}}%
}}

\def\horsize#1#2{%
\settowidth{\tempdimen}{$#2$}%
#1=\tempdimen \divide #1 by\unitlength }

\def\vertsize#1#2{%
\settoheight{\tempdimen}{$#2$}%
#1=\tempdimen
\settodepth{\tempdimen}{$#2$}%
\advance #1 by\tempdimen \divide #1 by\unitlength }

\def\putvector(#1,#2)(#3,#4)#5#6{{%
\ifnum3<\arrowtype \putdashvector(#1,#2)(#3,#4)#5\arrowtype \else
\ifnum\arrowtype<-3 \putdashvector(#1,#2)(#3,#4)#5\arrowtype \else
\xpos=#1 \ypos=#2 \run=#3 \rise=#4 \arrowlength=#5 \ifnum
\arrowtype<0
    \ifnum \run=0
        \advance \ypos by-\arrowlength
    \else
        \tempcounta \arrowlength
        \multiply \tempcounta by\rise
        \divide \tempcounta by\run
        \ifnum\run>0
            \advance \xpos by\arrowlength
            \advance \ypos by\tempcounta
        \else
            \advance \xpos by-\arrowlength
            \advance \ypos by-\tempcounta
        \fi
    \fi
    \multiply \arrowtype by-1
    \multiply \rise by-1
    \multiply \run by-1
\fi \ifcase \arrowtype
\or \put(\xpos,\ypos){\vector(\run,\rise){\arrowlength}}%
\or \put(\xpos,\ypos){\mvector(\run,\rise)\arrowlength}%
\or \put(\xpos,\ypos){\evector(\run,\rise){\arrowlength}}%
\fi\fi\fi }}

\def\putsplitvector(#1,#2)#3#4{
\xpos #1 \ypos #2 \arrowtype #4 \halflength #3 \arrowlength #3
\gap 140 \advance \halflength by-\gap \divide \halflength by2
\ifnum\arrowtype>0
   \ifcase \arrowtype
   \or \put(\xpos,\ypos){\line(0,-1){\halflength}}%
       \advance\ypos by-\halflength
       \advance\ypos by-\gap
       \put(\xpos,\ypos){\vector(0,-1){\halflength}}%
   \or \put(\xpos,\ypos){\line(0,-1)\halflength}%
       \put(\xpos,\ypos){\vector(0,-1)3}%
       \advance\ypos by-\halflength
       \advance\ypos by-\gap
       \put(\xpos,\ypos){\vector(0,-1){\halflength}}%
   \or \put(\xpos,\ypos){\line(0,-1)\halflength}%
       \advance\ypos by-\halflength
       \advance\ypos by-\gap
       \put(\xpos,\ypos){\evector(0,-1){\halflength}}%
   \fi
\else \arrowtype=-\arrowtype
   \ifcase\arrowtype
   \or \advance \ypos by-\arrowlength
       \put(\xpos,\ypos){\line(0,1){\halflength}}%
       \advance\ypos by\halflength
       \advance\ypos by\gap
       \put(\xpos,\ypos){\vector(0,1){\halflength}}%
   \or \advance \ypos by-\arrowlength
       \put(\xpos,\ypos){\line(0,1)\halflength}%
       \put(\xpos,\ypos){\vector(0,1)3}%
       \advance\ypos by\halflength
       \advance\ypos by\gap
       \put(\xpos,\ypos){\vector(0,1){\halflength}}%
   \or \advance \ypos by-\arrowlength
       \put(\xpos,\ypos){\line(0,1)\halflength}%
       \advance\ypos by\halflength
       \advance\ypos by\gap
       \put(\xpos,\ypos){\evector(0,1){\halflength}}%
   \fi
\fi }

\def\putmorphism(#1)(#2,#3)[#4`#5`#6]#7#8#9{{%
\run #2 \rise #3 \ifnum\rise=0
  \puthmorphism(#1)[#4`#5`#6]{#7}{#8}#9%
\else\ifnum\run=0
  \putvmorphism(#1)[#4`#5`#6]{#7}{#8}#9%
\else
\setpos(#1)%
\arrowlength #7 \arrowtype #8 \ifnum\run=0 \else\ifnum\rise=0
\else \ifnum\run>0
    \coefa=1
\else
   \coefa=-1
\fi \ifnum\arrowtype>0
   \coefb=0
   \coefc=-1
\else
   \coefb=\coefa
   \coefc=1
   \arrowtype=-\arrowtype
\fi \width=2 \multiply \width by\run \divide \width by\rise
\ifnum \width<0  \width=-\width\fi \advance\width by60 \if l#9
\width=-\width\fi
\putbox(\xpos,\ypos){#4}
{\multiply \coefa by\arrowlength
\advance\xpos by\coefa \multiply \coefa by\rise \divide \coefa
by\run \advance \ypos by\coefa
\putbox(\xpos,\ypos){#5} }%
{\multiply \coefa by\arrowlength
\divide \coefa by2 \advance \xpos by\coefa \advance \xpos by\width
\multiply \coefa by\rise \divide \coefa by\run \advance \ypos
by\coefa
\if l#9%
   \putrbox(\xpos,\ypos){#6}%
\else\if r#9%
   \putlbox(\xpos,\ypos){#6}%
\fi\fi }%
{\multiply \rise by-\coefc
\multiply \run by-\coefc \multiply \coefb by\arrowlength \advance
\xpos by\coefb \multiply \coefb by\rise \divide \coefb by\run
\advance \ypos by\coefb \multiply \coefc by70 \advance \ypos
by\coefc \multiply \coefc by\run \divide \coefc by\rise \advance
\xpos by\coefc \multiply \coefa by140 \multiply \coefa by\run
\divide \coefa by\rise \advance \arrowlength by\coefa
\ifcase\arrowtype
\or \put(\xpos,\ypos){\vector(\run,\rise){\arrowlength}}%
\or \put(\xpos,\ypos){\mvector(\run,\rise){\arrowlength}}%
\or \put(\xpos,\ypos){\evector(\run,\rise){\arrowlength}}%
\fi}\fi\fi\fi\fi}}

\newcount\numbdashes \newcount\lengthdash \newcount\increment

\def\howmanydashes{
\numbdashes=\arrowlength \lengthdash=40 \divide\numbdashes by
\lengthdash \lengthdash=\arrowlength \divide\lengthdash by
\numbdashes
\increment=\lengthdash \multiply\lengthdash by 3
\divide\lengthdash by 5 }

\def\putdashvector(#1)(#2,#3)#4#5{%
\ifnum#3=0 \putdashhvector(#1){#4}#5 \else \ifnum#2=0
\putdashvvector(#1){#4}#5\fi\fi}

\def\putdashhvector(#1,#2)#3#4{{%
\arrowlength=#3 \howmanydashes
\multiput(#1,#2)(\increment,0){\numbdashes}%
{\vrule height .4pt width \lengthdash\unitlength} \arrowtype=#4
\xpos=#1 \ifnum\arrowtype<0 \advance\arrowtype by 7 \fi
\ifcase\arrowtype \or \advance\xpos by 10
    \put(\xpos,#2){\vector(-1,0){\lengthdash}}
    \advance\xpos by 40
    \put(\xpos,#2){\vector(-1,0){\lengthdash}}
\or \advance \xpos by 10
    \put(\xpos,#2){\vector(-1,0){\lengthdash}}
    \advance\xpos by  \arrowlength
    \advance\xpos by  -50
    \put(\xpos,#2){\vector(-1,0){\lengthdash}}
\or \advance\xpos by 10
    \put(\xpos,#2){\vector(-1,0){\lengthdash}}
\or \advance\xpos by \arrowlength
    \advance\xpos by -\lengthdash
    \put(\xpos,#2){\vector(1,0){\lengthdash}}
\or {\advance\xpos by 10
    \put(\xpos,#2){\vector(1,0){\lengthdash}}}
    \advance\xpos by \arrowlength
    \advance\xpos by -\lengthdash
    \put(\xpos,#2){\vector(1,0){\lengthdash}}
\or \advance\xpos by \arrowlength
    \advance\xpos by -\lengthdash
    \put(\xpos,#2){\vector(1,0){\lengthdash}}
    \advance\xpos by -40
    \put(\xpos,#2){\vector(1,0){\lengthdash}}
   \fi
}}

\def\putdashvvector(#1,#2)#3#4{{%
\arrowlength=#3 \howmanydashes \ypos=#2 \advance\ypos by
-\arrowlength
\multiput(#1,#2)(0,\increment){\numbdashes}%
    {\vrule width .4pt height \lengthdash\unitlength}
\arrowtype=#4 \ypos=#2 \ifnum\arrowtype<0 \advance\arrowtype by 7
\fi \ifcase\arrowtype \or \advance\ypos by \arrowlength
\advance\ypos by -40
    \put(#1,\ypos){\vector(0,1){\lengthdash}}
    \advance\ypos by -40
    \put(#1,\ypos){\vector(0,1){\lengthdash}}
\or \advance\ypos by 10
    \put(#1,\ypos){\vector(0,1){\lengthdash}}
    \advance\ypos by \arrowlength \advance\ypos by -40
    \put(#1,\ypos){\vector(0,1){\lengthdash}}
\or \advance\ypos by \arrowlength \advance\ypos by -40
    \put(#1,\ypos){\vector(0,1){\lengthdash}}
\or \advance\ypos by 10
    \put(#1,\ypos){\vector(0,-1){\lengthdash}}
\or \advance\ypos by 10
    \put(#1,\ypos){\vector(0,-1){\lengthdash}}
    \advance\ypos by \arrowlength \advance\ypos by -40
    \put(#1,\ypos){\vector(0,-1){\lengthdash}}
\or \advance\ypos by 10
    \put(#1,\ypos){\vector(0,-1){\lengthdash}}
    \advance\ypos by 40
    \put(#1,\ypos){\vector(0,-1){\lengthdash}}
\fi }}

\def\puthmorphism(#1,#2)[#3`#4`#5]#6#7#8{{%
\xpos #1 \ypos #2 \width #6 \arrowlength #6 \arrowtype=#7
\putbox(\xpos,\ypos){#3\vphantom{#4}}%
{\advance \xpos by\arrowlength
\putbox(\xpos,\ypos){\vphantom{#3}#4}}%
\horsize{\tempcounta}{#3}%
\horsize{\tempcountb}{#4}%
\divide \tempcounta by2 \divide \tempcountb by2 \advance
\tempcounta by30 \advance \tempcountb by30 \advance \xpos
by\tempcounta \advance \arrowlength by-\tempcounta \advance
\arrowlength by-\tempcountb
\putvector(\xpos,\ypos)(1,0)\arrowlength\arrowtype \divide
\arrowlength by2 \advance \xpos by\arrowlength
\vertsize{\tempcounta}{#5}%
\divide\tempcounta by2 \advance \tempcounta by20
\if a#8 %
   \advance \ypos by\tempcounta
   \putbox(\xpos,\ypos){#5}%
\else
   \advance \ypos by-\tempcounta
   \putbox(\xpos,\ypos){#5}%
\fi}}

\def\putvmorphism(#1,#2)[#3`#4`#5]#6#7#8{{%
\xpos #1 \ypos #2 \arrowlength #6 \arrowtype #7
\settowidth{\xlen}{$#5$}%
\putbox(\xpos,\ypos){#3}%
{\advance \ypos by-\arrowlength
\putbox(\xpos,\ypos){#4}}%
{\advance\arrowlength by-140 \advance \ypos by-70 \ifdim\xlen>0pt
   \if m#8%
      \putsplitvector(\xpos,\ypos)\arrowlength\arrowtype
   \else
   \putvector(\xpos,\ypos)(0,-1)\arrowlength\arrowtype
   \fi
\else
   \putvector(\xpos,\ypos)(0,-1)\arrowlength\arrowtype
\fi}%
\ifdim\xlen>0pt
   \divide \arrowlength by2
   \advance\ypos by-\arrowlength
   \if l#8%
      \advance \xpos by-40
      \putrbox(\xpos,\ypos){#5}%
   \else\if r#8%
      \advance \xpos by40
      \putlbox(\xpos,\ypos){#5}%
   \else
      \putbox(\xpos,\ypos){#5}%
   \fi\fi
\fi }}

\def\putsquarep<#1>(#2)[#3;#4`#5`#6`#7]{{%
\setsqparms[#1]%
\setpos(#2)%
\settokens`#3`%
\puthmorphism(\xpos,\ypos)[\tokenc`\tokend`{#7}]{\width}{\arrowtyped}b%
\advance\ypos by \height
\puthmorphism(\xpos,\ypos)[\tokena`\tokenb`{#4}]{\width}{\arrowtypea}a%
\putvmorphism(\xpos,\ypos)[``{#5}]{\height}{\arrowtypeb}l%
\advance\xpos by \width
\putvmorphism(\xpos,\ypos)[``{#6}]{\height}{\arrowtypec}r%
}}

\def\putsquare{\@ifnextchar <{\putsquarep}{\putsquarep%
   <\arrowtypea`\arrowtypeb`\arrowtypec`\arrowtyped;\width`\height>}}
\def\square{\@ifnextchar< {\squarep}{\squarep
   <\arrowtypea`\arrowtypeb`\arrowtypec`\arrowtyped;\width`\height>}}
\def\squarep<#1>[#2`#3`#4`#5;#6`#7`#8`#9]{{
\setsqparms[#1]
\diagram
\putsquarep<\arrowtypea`\arrowtypeb`\arrowtypec`
\arrowtyped;\width`\height>
(0,0)[#2`#3`#4`{#5};#6`#7`#8`{#9}]
\enddiagram
}}                                                 
\def\putptrianglep<#1>(#2,#3)[#4`#5`#6;#7`#8`#9]{{%
\settriparms[#1]%
\xpos=#2 \ypos=#3 \advance\ypos by \height
\puthmorphism(\xpos,\ypos)[#4`#5`{#7}]{\height}{\arrowtypea}a%
\putvmorphism(\xpos,\ypos)[`#6`{#8}]{\height}{\arrowtypeb}l%
\advance\xpos by\height
\putmorphism(\xpos,\ypos)(-1,-1)[``{#9}]{\height}{\arrowtypec}r%
}}

\def\putptriangle{\@ifnextchar <{\putptrianglep}{\putptrianglep
   <\arrowtypea`\arrowtypeb`\arrowtypec;\height>}}
\def\ptriangle{\@ifnextchar <{\ptrianglep}{\ptrianglep
   <\arrowtypea`\arrowtypeb`\arrowtypec;\height>}}
\def\ptrianglep<#1>[#2`#3`#4;#5`#6`#7]{{
\settriparms[#1]
\diagram
\putptrianglep<\arrowtypea`\arrowtypeb`
\arrowtypec;\height>
(0,0)[#2`#3`#4;#5`#6`{#7}]
\enddiagram
}}                                            

\def\putqtrianglep<#1>(#2,#3)[#4`#5`#6;#7`#8`#9]{{%
\settriparms[#1]%
\xpos=#2 \ypos=#3 \advance\ypos by\height
\puthmorphism(\xpos,\ypos)[#4`#5`{#7}]{\height}{\arrowtypea}a%
\putmorphism(\xpos,\ypos)(1,-1)[``{#8}]{\height}{\arrowtypeb}l%
\advance\xpos by\height
\putvmorphism(\xpos,\ypos)[`#6`{#9}]{\height}{\arrowtypec}r%
}}

\def\putqtriangle{\@ifnextchar <{\putqtrianglep}{\putqtrianglep
   <\arrowtypea`\arrowtypeb`\arrowtypec;\height>}}
\def\qtriangle{\@ifnextchar <{\qtrianglep}{\qtrianglep
   <\arrowtypea`\arrowtypeb`\arrowtypec;\height>}}
\def\qtrianglep<#1>[#2`#3`#4;#5`#6`#7]{{
\settriparms[#1]
\width=\height                                
\diagram
\putqtrianglep<\arrowtypea`\arrowtypeb`
\arrowtypec;\height>
(0,0)[#2`#3`#4;#5`#6`{#7}]
\enddiagram
}}

\def\putdtrianglep<#1>(#2,#3)[#4`#5`#6;#7`#8`#9]{{%
\settriparms[#1]%
\xpos=#2 \ypos=#3
\puthmorphism(\xpos,\ypos)[#5`#6`{#9}]{\height}{\arrowtypec}b%
\advance\xpos by \height \advance\ypos by\height
\putmorphism(\xpos,\ypos)(-1,-1)[``{#7}]{\height}{\arrowtypea}l%
\putvmorphism(\xpos,\ypos)[#4``{#8}]{\height}{\arrowtypeb}r%
}}

\def\putdtriangle{\@ifnextchar <{\putdtrianglep}{\putdtrianglep
   <\arrowtypea`\arrowtypeb`\arrowtypec;\height>}}
\def\dtriangle{\@ifnextchar <{\dtrianglep}{\dtrianglep
   <\arrowtypea`\arrowtypeb`\arrowtypec;\height>}}
\def\dtrianglep<#1>[#2`#3`#4;#5`#6`#7]{{
\settriparms[#1]
\width=\height                                
\diagram
\putdtrianglep<\arrowtypea`\arrowtypeb`
\arrowtypec;\height>
(0,0)[#2`#3`#4;#5`#6`{#7}]
\enddiagram
}}

\def\putbtrianglep<#1>(#2,#3)[#4`#5`#6;#7`#8`#9]{{%
\settriparms[#1]%
\xpos=#2 \ypos=#3
\puthmorphism(\xpos,\ypos)[#5`#6`{#9}]{\height}{\arrowtypec}b%
\advance\ypos by\height
\putmorphism(\xpos,\ypos)(1,-1)[``{#8}]{\height}{\arrowtypeb}r%
\putvmorphism(\xpos,\ypos)[#4``{#7}]{\height}{\arrowtypea}l%
}}

\def\putbtriangle{\@ifnextchar <{\putbtrianglep}{\putbtrianglep
   <\arrowtypea`\arrowtypeb`\arrowtypec;\height>}}
\def\btriangle{\@ifnextchar <{\btrianglep}{\btrianglep
   <\arrowtypea`\arrowtypeb`\arrowtypec;\height>}}
\def\btrianglep<#1>[#2`#3`#4;#5`#6`#7]{{
\settriparms[#1]
\width=\height                               
\diagram
\putbtrianglep<\arrowtypea`\arrowtypeb`
\arrowtypec;\height>
(0,0)[#2`#3`#4;#5`#6`{#7}]
\enddiagram
}}

\def\putAtrianglep<#1>(#2,#3)[#4`#5`#6;#7`#8`#9]{{%
\settriparms[#1]%
\xpos=#2 \ypos=#3 {\multiply \height by2
\puthmorphism(\xpos,\ypos)[#5`#6`{#9}]{\height}{\arrowtypec}b}%
\advance\xpos by\height \advance\ypos by\height
\putmorphism(\xpos,\ypos)(-1,-1)[#4``{#7}]{\height}{\arrowtypea}l%
\putmorphism(\xpos,\ypos)(1,-1)[``{#8}]{\height}{\arrowtypeb}r%
}}

\def\putAtriangle{\@ifnextchar <{\putAtrianglep}{\putAtrianglep
   <\arrowtypea`\arrowtypeb`\arrowtypec;\height>}}
\def\Atriangle{\@ifnextchar <{\Atrianglep}{\Atrianglep
   <\arrowtypea`\arrowtypeb`\arrowtypec;\height>}}
\def\Atrianglep<#1>[#2`#3`#4;#5`#6`#7]{{
\settriparms[#1]
\width=\height                                     
\diagram
\putAtrianglep<\arrowtypea`\arrowtypeb`
\arrowtypec;\height>
(0,0)[#2`#3`#4;#5`#6`{#7}]
\enddiagram
}}

\def\putAtrianglepairp<#1>(#2)[#3;#4`#5`#6`#7`#8]{{%
\settripairparms[#1]%
\setpos(#2)%
\settokens`#3`%
\puthmorphism(\xpos,\ypos)[\tokenb`\tokenc`{#7}]{\height}{\arrowtyped}b%
\advance\xpos by\height
\puthmorphism(\xpos,\ypos)[\phantom{\tokenc}`\tokend`{#8}]%
{\height}{\arrowtypee}b%
\advance\ypos by\height
\putmorphism(\xpos,\ypos)(-1,-1)[\tokena``{#4}]{\height}{\arrowtypea}l%
\putvmorphism(\xpos,\ypos)[``{#5}]{\height}{\arrowtypeb}m%
\putmorphism(\xpos,\ypos)(1,-1)[``{#6}]{\height}{\arrowtypec}r%
}}

\def\putAtrianglepair{\@ifnextchar <{\putAtrianglepairp}{\putAtrianglepairp%
   <\arrowtypea`\arrowtypeb`\arrowtypec`\arrowtyped`\arrowtypee;\height>}}
\def\Atrianglepair{\@ifnextchar <{\Atrianglepairp}{\Atrianglepairp%
   <\arrowtypea`\arrowtypeb`\arrowtypec`\arrowtyped`\arrowtypee;\height>}}

\def\Atrianglepairp<#1>[#2;#3`#4`#5`#6`#7]{{
\settripairparms[#1]
\settokens`#2`
\width=\height                                
\diagram
\putAtrianglepairp                            
<\arrowtypea`\arrowtypeb`\arrowtypec`
\arrowtyped`\arrowtypee;\height>
(0,0)[{#2};#3`#4`#5`#6`{#7}]
\enddiagram
}}

\def\putVtrianglep<#1>(#2,#3)[#4`#5`#6;#7`#8`#9]{{%
\settriparms[#1]%
\xpos=#2 \ypos=#3 \advance\ypos by\height {\multiply\height by2
\puthmorphism(\xpos,\ypos)[#4`#5`{#7}]{\height}{\arrowtypea}a}%
\putmorphism(\xpos,\ypos)(1,-1)[`#6`{#8}]{\height}{\arrowtypeb}l%
\advance\xpos by\height \advance\xpos by\height
\putmorphism(\xpos,\ypos)(-1,-1)[``{#9}]{\height}{\arrowtypec}r%
}}

\def\putVtriangle{\@ifnextchar <{\putVtrianglep}{\putVtrianglep
   <\arrowtypea`\arrowtypeb`\arrowtypec;\height>}}
\def\Vtriangle{\@ifnextchar <{\Vtrianglep}{\Vtrianglep
   <\arrowtypea`\arrowtypeb`\arrowtypec;\height>}}
\def\Vtrianglep<#1>[#2`#3`#4;#5`#6`#7]{{
\settriparms[#1]
\width=\height                                 
\diagram
\putVtrianglep<\arrowtypea`\arrowtypeb`
\arrowtypec;\height>
(0,0)[#2`#3`#4;#5`#6`{#7}]
\enddiagram
}}

\def\putVtrianglepairp<#1>(#2)[#3;#4`#5`#6`#7`#8]{{
\settripairparms[#1]%
\setpos(#2)%
\settokens`#3`%
\advance\ypos by\height
\putmorphism(\xpos,\ypos)(1,-1)[`\tokend`{#6}]{\height}{\arrowtypec}l%
\puthmorphism(\xpos,\ypos)[\tokena`\tokenb`{#4}]{\height}{\arrowtypea}a%
\advance\xpos by\height
\puthmorphism(\xpos,\ypos)[\phantom{\tokenb}`\tokenc`{#5}]%
{\height}{\arrowtypeb}a%
\putvmorphism(\xpos,\ypos)[``{#7}]{\height}{\arrowtyped}m%
\advance\xpos by\height
\putmorphism(\xpos,\ypos)(-1,-1)[``{#8}]{\height}{\arrowtypee}r%
}}

\def\putVtrianglepair{\@ifnextchar <{\putVtrianglepairp}{\putVtrianglepairp%
    <\arrowtypea`\arrowtypeb`\arrowtypec`\arrowtyped`\arrowtypee;\height>}}
\def\Vtrianglepair{\@ifnextchar <{\Vtrianglepairp}{\Vtrianglepairp%
    <\arrowtypea`\arrowtypeb`\arrowtypec`\arrowtyped`\arrowtypee;\height>}}
\def\Vtrianglepairp<#1>[#2;#3`#4`#5`#6`#7]{{
\settripairparms[#1]
\settokens`#2`
\diagram
\putVtrianglepairp                             
<\arrowtypea`\arrowtypeb`\arrowtypec`
\arrowtyped`\arrowtypee;\height>
(0,0)[{#2};#3`#4`#5`#6`{#7}]
\enddiagram
}}

\def\putCtrianglep<#1>(#2,#3)[#4`#5`#6;#7`#8`#9]{{%
\settriparms[#1]%
\xpos=#2 \ypos=#3 \advance\ypos by\height
\putmorphism(\xpos,\ypos)(1,-1)[``{#9}]{\height}{\arrowtypec}l%
\advance\xpos by\height \advance\ypos by\height
\putmorphism(\xpos,\ypos)(-1,-1)[#4`#5`{#7}]{\height}{\arrowtypea}l%
{\multiply\height by 2
\putvmorphism(\xpos,\ypos)[`#6`{#8}]{\height}{\arrowtypeb}r}%
}}

\def\putCtriangle{\@ifnextchar <{\putCtrianglep}{\putCtrianglep
    <\arrowtypea`\arrowtypeb`\arrowtypec;\height>}}
\def\Ctriangle{\@ifnextchar <{\Ctrianglep}{\Ctrianglep
    <\arrowtypea`\arrowtypeb`\arrowtypec;\height>}}
\def\Ctrianglep<#1>[#2`#3`#4;#5`#6`#7]{{
\settriparms[#1]
\width=\height                               
\diagram
\putCtrianglep<\arrowtypea`\arrowtypeb`
\arrowtypec;\height>
(0,0)[#2`#3`#4;#5`#6`{#7}]
\enddiagram
}}                                           
\def\putDtrianglep<#1>(#2,#3)[#4`#5`#6;#7`#8`#9]{{%
\settriparms[#1]%
\xpos=#2 \ypos=#3 \advance\xpos by\height \advance\ypos by\height
\putmorphism(\xpos,\ypos)(-1,-1)[``{#9}]{\height}{\arrowtypec}r%
\advance\xpos by-\height \advance\ypos by\height
\putmorphism(\xpos,\ypos)(1,-1)[`#5`{#8}]{\height}{\arrowtypeb}r%
{\multiply\height by 2
\putvmorphism(\xpos,\ypos)[#4`#6`{#7}]{\height}{\arrowtypea}l}%
}}

\def\putDtriangle{\@ifnextchar <{\putDtrianglep}{\putDtrianglep
    <\arrowtypea`\arrowtypeb`\arrowtypec;\height>}}
\def\Dtriangle{\@ifnextchar <{\Dtrianglep}{\Dtrianglep
   <\arrowtypea`\arrowtypeb`\arrowtypec;\height>}}
\def\Dtrianglep<#1>[#2`#3`#4;#5`#6`#7]{{
\settriparms[#1]
\width=\height                              
\diagram
\putDtrianglep<\arrowtypea`\arrowtypeb`
\arrowtypec;\height>
(0,0)[#2`#3`#4;#5`#6`{#7}]
\enddiagram
}}                                          
\def\setrecparms[#1`#2]{\width=#1 \height=#2}%

\def\recursep<#1`#2>[#3;#4`#5`#6`#7`#8]{{\m@th
\width=#1 \height=#2 \settokens`#3`
\settowidth{\tempdimen}{$\tokena$} \ifdim\tempdimen=0pt
  \savebox{\tempboxa}{\hbox{$\tokenb$}}%
  \savebox{\tempboxb}{\hbox{$\tokend$}}%
  \savebox{\tempboxc}{\hbox{$#6$}}%
\else
  \savebox{\tempboxa}{\hbox{$\hbox{$\tokena$}\times\hbox{$\tokenb$}$}}%
  \savebox{\tempboxb}{\hbox{$\hbox{$\tokena$}\times\hbox{$\tokend$}$}}%
  \savebox{\tempboxc}{\hbox{$\hbox{$\tokena$}\times\hbox{$#6$}$}}%
\fi \ypos=\height \divide\ypos by 2 \xpos=\ypos \advance\xpos by
\width \bfig
\putCtrianglep<-1`1`1;\ypos>(0,0)[`\tokenc`;#5`#6`{#7}]%
\puthmorphism(\ypos,0)[\tokend`\usebox{\tempboxb}`{#8}]{\width}{-1}b%
\puthmorphism(\ypos,\height)[\tokenb`\usebox{\tempboxa}`{#4}]{\width}{-1}a%
\advance\ypos by \width
\putvmorphism(\ypos,\height)[``\usebox{\tempboxc}]{\height}1r%
\efig }}

\def\recurse{\@ifnextchar <{\recursep}{\recursep<\width`\height>}}

\def\puttwohmorphisms(#1,#2)[#3`#4;#5`#6]#7#8#9{{%
\puthmorphism(#1,#2)[#3`#4`]{#7}0a \ypos=#2 \advance\ypos by 20
\puthmorphism(#1,\ypos)[\phantom{#3}`\phantom{#4}`#5]{#7}{#8}a
\advance\ypos by -40
\puthmorphism(#1,\ypos)[\phantom{#3}`\phantom{#4}`#6]{#7}{#9}b }}

\def\puttwovmorphisms(#1,#2)[#3`#4;#5`#6]#7#8#9{{%
\putvmorphism(#1,#2)[#3`#4`]{#7}0a \xpos=#1 \advance\xpos by -20
\putvmorphism(\xpos,#2)[\phantom{#3}`\phantom{#4}`#5]{#7}{#8}l
\advance\xpos by 40
\putvmorphism(\xpos,#2)[\phantom{#3}`\phantom{#4}`#6]{#7}{#9}r }}

\def\puthcoequalizer(#1)[#2`#3`#4;#5`#6`#7]#8#9{{%
\setpos(#1)%
\puttwohmorphisms(\xpos,\ypos)[#2`#3;#5`#6]{#8}11%
\advance\xpos by #8
\puthmorphism(\xpos,\ypos)[\phantom{#3}`#4`#7]{#8}1{#9} }}

\def\putvcoequalizer(#1)[#2`#3`#4;#5`#6`#7]#8#9{{%
\setpos(#1)%
\puttwovmorphisms(\xpos,\ypos)[#2`#3;#5`#6]{#8}11%
\advance\ypos by -#8
\putvmorphism(\xpos,\ypos)[\phantom{#3}`#4`#7]{#8}1{#9} }}

\def\putthreehmorphisms(#1)[#2`#3;#4`#5`#6]#7(#8)#9{{%
\setpos(#1) \settypes(#8)
\if a#9 %
     \vertsize{\tempcounta}{#5}%
     \vertsize{\tempcountb}{#6}%
     \ifnum \tempcounta<\tempcountb \tempcounta=\tempcountb \fi
\else
     \vertsize{\tempcounta}{#4}%
     \vertsize{\tempcountb}{#5}%
     \ifnum \tempcounta<\tempcountb \tempcounta=\tempcountb \fi
\fi \advance \tempcounta by 60
\puthmorphism(\xpos,\ypos)[#2`#3`#5]{#7}{\arrowtypeb}{#9}
\advance\ypos by \tempcounta
\puthmorphism(\xpos,\ypos)[\phantom{#2}`\phantom{#3}`#4]{#7}{\arrowtypea}{#9}
\advance\ypos by -\tempcounta \advance\ypos by -\tempcounta
\puthmorphism(\xpos,\ypos)[\phantom{#2}`\phantom{#3}`#6]{#7}{\arrowtypec}{#9}
}}

\def\setarrowtoks[#1`#2`#3`#4`#5`#6]{%
\def\toka{#1}
\def\tokb{#2}
\def\tokc{#3}
\def\tokd{#4}
\def\toke{#5}
\def\tokf{#6}
}
\def\hex{\@ifnextchar <{\hexp}{\hexp<1000`400>}}
\def\hexp<#1`#2>[#3`#4`#5`#6`#7`#8;#9]{%
\setarrowtoks[#9] \yext=#2 \advance \yext by #2 \xext=#1
\advance\xext by \yext \bfig
\putCtriangle<-1`0`1;#2>(0,0)[`#5`;\tokb``\tokd] \xext=#1
\yext=#2 \advance \yext by #2
\putsquare<1`0`0`1;\xext`\yext>(#2,0)[#3`#4`#7`#8;\toka```\tokf]
\advance \xext by #2
\putDtriangle<0`1`-1;#2>(\xext,0)[`#6`;`\tokc`\toke] \efig }

\makeatother

\begin{document}

\title{A Theoretical Model of Chaotic Attractor\\ in Tumor Growth and Metastasis}
\author{Tijana T. Ivancevic, Murk J. Bottema and Lakhmi C. Jain}
\date{}
\maketitle

\begin{abstract} This paper proposes a novel chaotic
reaction-diffusion model of cellular tumor growth and metastasis.
The model is based on the  multiscale diffusion cancer-invasion
model (MDCM) and formulated by introducing strong nonlinear
coupling into the MDCM. The new model exhibits temporal chaotic
behavior (which resembles the classical Lorenz strange attractor)
and yet retains all the characteristics of the MDCM diffusion
model. It mathematically describes both the processes of
carcinogenesis and metastasis, as well as the sensitive dependence
of cancer evolution on initial conditions and parameters. On the
basis of this chaotic tumor-growth model, a generic concept of
carcinogenesis and metastasis is formulated. \bigbreak

\end{abstract}

\section{Introduction}

Cancer is one of the main causes of morbidity and mortality in the
world. There are several different stages in the growth of a tumor
before it becomes so large that it causes the patient to die or
reduces permanently their quality of life. Developed countries are
investing large sums of money into cancer research in order to
find cures and improve existing treatments. In comparison to
molecular biology, cell biology, and drug delivery research,
mathematics has so far contributed relatively little to the area
\cite{Nature}.

A number of mathematical models of avascular (solid) tumor growth
were reviewed in \cite{Roose}. These were  generally divided into
continuum cell population models described by diffusion partial
differential equations (PDEs) of continuum mechanics
\cite{GaneshSprBig,GCompl} combined with chemical kinetics, and
discrete cell population models described by ordinary differential
equations (ODEs).

On the other hand, in many biological systems it is possible to
\emph{empirically} demonstrate the presence of \emph{attractors}
that operate starting from different initial conditions
\cite{TacaNODY}. Some of these attractors are points, some are
closed curves, while the others have non--integer, fractal
dimension and are termed ``strange attractors" \cite{horizons}. It
has been proposed that a prerequisite for proper simulating tumor
growth by computer is to establish whether typical tumor growth
patterns are fractal. The fractal dimension of tumor outlines was
empirically determined using the \emph{box-counting} method
\cite{frac}. In particular, fractal analysis of a breast carcinoma
was performed using a \emph{morphometric method}, which is the
box-counting method applied to the mammogram as well as to the
histologic section of a breast carcinoma \cite{frac2}.

If tumor growth is chaotic, this could explain the unreliability
of treatment and prediction of tumor evolution. More importantly,
if chaos is established, this could be used to adjust strategies
for fighting cancer. Treatment could include some form of
\emph{chaos control} and/or \emph{anti-control}.\footnote{The
chaotic behavior of a system may be artificially weakened or
suppressed if it is undesirable. This concept is known as
\emph{control of chaos}. The first and the most important method
of chaos control is the so--called OGY--method, developed by
\cite{OGY}. However, in recent years, a non-traditional concept of
anti-control of chaos has emerged. Here, the non-chaotic dynamical
system is transformed into a chaotic one by small controlled
perturbation so that useful properties of a chaotic system can be
utilized \cite{StrAttr,Complexity}.}

In this paper, a plausible chaotic diffusion model of tumor
growth and metastasis is presented. The approach is a combination of
theoretical modelling and empirical search.

\section{A multiscale diffusion cancer-invasion model}

Recently, a multiscale diffusion cancer-invasion model (MDCM) was
presented in
\cite{Anderson98,Anderson00,Anderson05,Anderson-Cell,Anderson06,AndersonMBE,AndersonJTB,Anderson08,Chaplain08},
which considers cellular and microenvironmental factors
simultaneously and interactively. The model was classified as
\emph{hybrid}, since a continuum deterministic model (based on a
system of reaction--diffusion chemotaxis equations) controls the
chemical and extracellular matrix (ECM) kinetics and a discrete
cellular automata-like model (based on a biased random-walk model)
controls the cell migration and interaction. The interactions of
the tumor cells, matrix--metalloproteinases (MMs),
matrix-degradative enzymes (MDEs) and oxygen are described by the
four coupled rate PDEs:
\begin{eqnarray}
\frac{\partial n}{\partial t} &=&D_{n}\nabla ^{2}n-\chi \nabla
\cdot
(n\nabla f),  \label{c1} \\
\frac{\partial f}{\partial t} &=&-\delta mf, \label{c2} \\
\frac{\partial m}{\partial t} &=&D_{m}\nabla ^{2}m+\mu n-\lambda m,
\label{c3} \\
\frac{\partial c}{\partial t} &=&D_{c}\nabla ^{2}c+\beta f-\gamma
n-\alpha c, \label{c4}
\end{eqnarray}
where $n$ denotes the tumor cell density, $f$ is the
MM--concentration, $m$ corresponds to the MDE--concentration, and
$c$ denotes the oxygen concentration. The four variables,
$n,m,f,c$, are all functions of the 3-dimensional
spatial variable $x$ and time
$t$. All equations represent diffusion except (\ref{c2}), which
shows only temporal evolution of the MM--concentration coupled to
the MDE--concentration. $D_{n}$ is the tumor cell coefficient, $D_{m}$
is the MDE coefficient and $D_{c}>0$ is the oxygen diffusion coefficient,
while $\chi ,\mu $, $\lambda $, $\delta $, $\alpha $, $\gamma $,
$\beta $ are positive constants. The other terms respectively
denote:

$\chi \nabla \cdot (n\nabla f)-$haptotaxis;

$\mu N-$production of MDE by tumor cell;

$\lambda m-$decay of MDE;

$\delta mf-$degradation of MM by MDE;

$\alpha c-$natural decay of oxygen;

$\gamma n-$oxygen uptake;  and

$\beta f-$production of oxygen by MM.

Because of its \emph{hybrid nature} (cells treated as discrete
entities and microenvironmental parameters treated as continuous
concentrations), the 4--dimensional (4D) model
(\ref{c1})--(\ref{c4}) can be directly linked to experimental
measurements of those cellular and microenvironmental parameters
recognized by cancer biologists as important in cancer invasion.
Furthermore, the fundamental unit of the model is the cell, and
the complex collective behavior of the tumor emerges as a
consequence of interactions between factors influencing the life
cycle and movement of individual cells
\cite{Anderson98,Anderson05,Anderson-Cell,AndersonJTB,Anderson08,Chaplain08}.

In order to use realistic parameter values, the system of rate
equations (\ref{c1})--(\ref{c4}) was non-dimensionalised.
The resulting 4D scaled system of rate PDEs
\cite{Anderson05} is given by
\begin{eqnarray}
\frac{\partial n}{\partial t} &=&d_{n}\nabla ^{2}n-\rho \nabla
\cdot
(n\nabla f),  \label{d1} \\
\frac{\partial f}{\partial t} &=&-\eta mf,  \label{d2} \\
\frac{\partial m}{\partial t} &=&d_{m}\nabla ^{2}m+\kappa n-\sigma
m,
\label{d3} \\
\frac{\partial c}{\partial t} &=&d_{c}\nabla ^{2}c+\nu f-\omega
n-\phi c. \label{d4}
\end{eqnarray}
In \cite{Anderson05} the values of the non-dimensional parameters
were given as:
\begin{eqnarray}
&&d_n = 0.0005, ~~d_m = 0.0005, ~~d_c = 0.5, ~~\rho = 0.01, ~~\eta
= 50, \notag
\\ &&\kappa = 1, ~~\sigma = 0, ~~\nu = 0.5, ~~\omega = 0.57, ~~\phi =
0.025. \label{par}
\end{eqnarray}

The 4D hybrid PDE--model (\ref{c1})--(\ref{c4}) can be seen as a
special case of a general \emph{multi--phase tumor growth PDE}
(\cite{Roose} equation (12)),
\begin{equation}
\frac{\partial \Phi _{i}}{\partial t}+\nabla \cdot (\mathbf{v}_{i}\Phi
_{i})=\nabla \cdot (D_{i}\Phi _{i})+\lambda _{i}(\Phi _{i},C_{i})-\mu
_{i}(\Phi _{i},C_{i}),  \label{multiPh}
\end{equation}
where for phase $i$, $\Phi _{i}$ is the volume fraction ($\sum_{i}\Phi
_{i}=1 $), $\mathbf{v}_{i}$ is the velocity, $D_{i}$ is the random motility
or diffusion, $\lambda _{i}(\Phi _{i},C_{i})$ is the chemical and phase
dependent production, and $\mu _{i}(\Phi _{i},C_{i})$ is the chemical and
phase dependent degradation/death. The multi-phase model (\ref{multiPh}) has
been derived from the basic conservation equations for the different
chemical species,
\[
\frac{\partial C_{i}}{\partial t}+\nabla \cdot \mathbf{N}_{i}=P_{i},
\]
where $C_{i}$ are the concentrations of the chemical species,
subindex $a$ for oxygen, $b$ for glucose, $c$ for lactate ion, $d$
for carbon dioxide, $e$ for bicarbonate ion, $f$ for chloride ion,
and $g$ for hydrogen ion concentration; $\mathbf{N}_{i}$ is the
flux of each of the chemical species inside the tumor spheroid;
and $P_{i}$ is the net rate of consumption/production of the
chemical species both by tumor cells and due to the chemical
reactions with other species.

In this paper we will search for a temporal 3D \emph{cancer
chaotic attractor}\footnote{An \emph{attractor} is defined as the
smallest set which cannot be itself decomposed into two or more
attractors with distinct \emph{basins of attraction} (this
restriction is necessary since a dynamical system may have
multiple attractors, each with its own basin of attraction). A
\emph{chaotic}, or \emph{strange attractor} is an attractor that
has zero measure in the embedding phase--space and has fractal
dimension. Trajectories within a strange attractor appear to skip
around randomly \cite{StrAttr,Complexity}. In particular, the
celebrated \emph{Lorenz attractor} is given by the set of ODEs
\cite{Lorenz,Sp}
\begin{equation}
\dot{x}=a(y-x),\qquad \dot{y}=bx-y-xz,\qquad \dot{z}=xy-cz,
\label{LorenzSys}
\end{equation}
where $x$, $y$ and $z$ are dynamical variables, constituting the 3D \emph{%
phase--space} of the \textit{Lorenz system}; and $a$, $b$ and $c$
are the parameters of the system. Originally, Lorenz used this
model to describe the unpredictable behavior of the weather, where
$x$ is the rate of convective overturning (convection is the
process by which heat is transferred by a moving fluid), $y$ is
the horizontal temperature overturning, and $z$ is the
vertical temperature overturning; the parameters are: $a\equiv P-$%
proportional to the \textit{Prandtl number} (ratio of the fluid
viscosity of
a substance to its thermal conductivity, usually set at $10$), $b\equiv R-$%
proportional to the Rayleigh number (difference in temperature
between the top and bottom of the system, usually set at $28$),
and $c\equiv K-$a number proportional to the physical proportions
of the region under consideration
(width to height ratio of the box which holds the system, usually set at $%
8/3 $). The Lorenz system (\ref{LorenzSys}) has the properties:
\begin{enumerate}
\item  \emph{Symmetry}: $(x,y,z)\rightarrow (-x,-y,z)$ for all
values of the parameters, and

\item  The $z-$axis $(x=y=0)$ is \emph{invariant} (i.e., all
trajectories that start on it also end on it).
\end{enumerate}}
`buried' within the 4D hybrid spatio-temporal model
(\ref{c1})--(\ref{c4}).

\section{A chaotic multi-scale cancer-invasion model}

From the non--dimensional spatio-temporal AC model
(\ref{d1})--(\ref{d4}), discretization was performed by neglecting
all the spatial derivatives resulting in the following simple 4D
temporal dynamical system.
\begin{eqnarray}
\dot{n} &=&0,  \label{cd1} \\
\dot{f} &=&-\eta mf,  \label{cd2} \\
\dot{m} &=&\kappa n-\sigma m,  \label{cd3} \\
\dot{c} &=&\nu f-\omega n-\phi c.  \label{cd4}
\end{eqnarray}

When simulated, the temporal system (\ref{cd1})--(\ref{cd4}) with
the set of parameters (\ref{par}) exhibits a virtually linear
temporal behavior with almost no coupling between the four
concentrations that have very different quantitative values (all
phase plots between the four concentrations, not shown here, are
virtually one-dimensional). To see if a modified version of the
system (\ref{cd1})--(\ref{cd4}) could lead to a chaotic
description of tumor growth, four new parameters, $\alpha$,
$\beta$, $\gamma$, and $\delta$ were introduced. The resulting
model is
\begin{eqnarray}
\dot{n} &=&0,  \label{md1} \\
\dot{f} &=&\alpha \eta (m-f),  \label{md2} \\
\dot{m} &=&\beta \kappa n+f(\gamma -c)-m,  \label{md3} \\
\dot{c} &=&\nu fm-\omega n-\delta \phi c.  \label{md4}
\end{eqnarray}

The introduction of the parameters $(\alpha,\beta,\gamma,\delta)$
was motivated by the fact that tumor cell shape (see Figures 6 and
7) represents a visual manifestation of an underlying balance of
forces and chemical reactions \cite{Olive}. Specifically, the
parameters represent the following quantities:

 $\alpha ~~ {\rm =} $~ tumor cell volume (proliferation/non-proliferation
fraction),

 $\beta ~~ {\rm =} $~ glucose level,

 $\gamma ~~ {\rm =} $~ number of tumor cells,

 $\delta ~~ {\rm =} $~ diffusion from the surface (saturation
 level).\smallskip

A tumor is composed of proliferating ($P$) and quiescent (or
non-proliferating) ($Q$) cells. Tumor cells shift from class $P $
to class $Q $ as the tumor grows in size \cite{Kozusko}. Model
dependence on the ratio of proliferation to non-proliferation is
introduced via the first parameter, $\alpha$. The discretization
of equation (\ref{d1}) leads to cell density being modelled as a
constant in equation (\ref{cd1}). Accordingly, cell density does
not play a role in the dynamics. In (\ref{md1})--(\ref{md4}) the
cell density is re-introduced into the dynamics via the cell
number, $\gamma$. The importance of introducing $\gamma$ also
appears in connection with the cyclin-dependent kinase (Cdk)
inhibitor p27, the level and activity of which increase in
response to cell density. Levels and activity of Cdk inhibitor p27
also increase with differentiation following loss of adhesion to
the ECM \cite{Chu}.

The ability to estimate the growth pattern of an individual tumor
cell type on the basis of morphological measurements should have
general applicability in cellular investigations, cell--growth
kinetics, cell transformation and morphogenesis \cite{Castro}.

Cell spreading alone is conducive to proliferation and increases
in DNA synthesis, indicating that cell morphology is a critical
determinant of cell function, at least in the presence of optimal
growth factors and extracellular matrix (ECM) binding
\cite{Ingber}. In many cells, the changes in morphology can
stimulate cell proliferation through integrin-mediated signaling,
indicating that cell shape may govern how individual cells will
respond to chemical signals \cite{Boudreau}.
\begin{figure}[ht]
\centerline{\includegraphics[width=9cm]{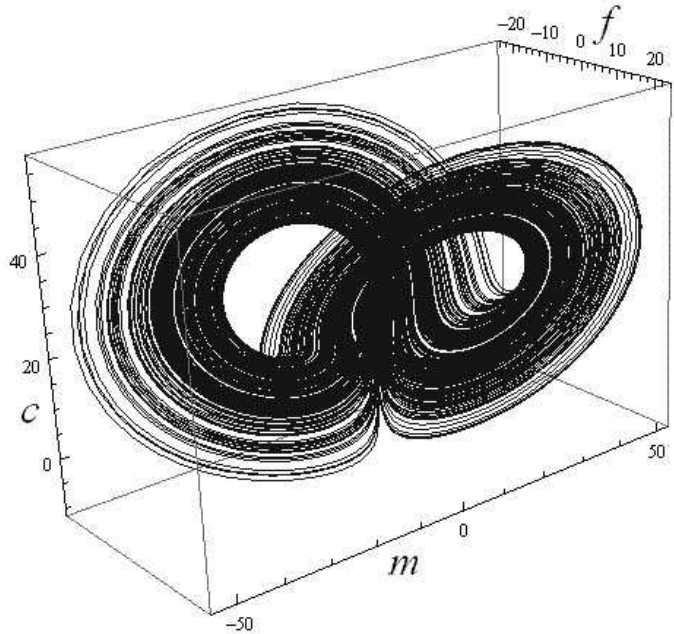}} \caption{A 3D
Lorenz-like chaotic attractor (see \cite{Lorenz,Sp}) from the
modified tumor growth model (\ref{md1})-(\ref{md4}). The attractor
effectively couples the MM--concentration $f$, the
MDE--concentration $m$, and the oxygen concentration $c$ in a
mask--like fashion.} \label{tuAtr1}
\end{figure}
\begin{figure}[ht]
\centerline{\includegraphics[width=8cm]{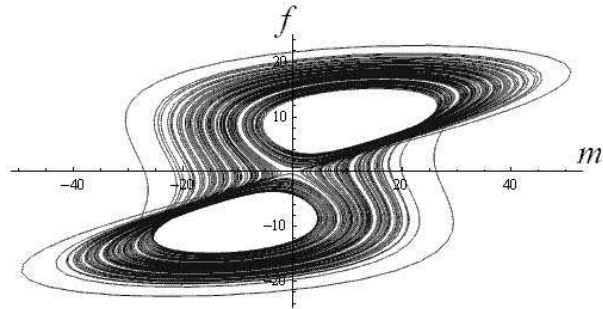}} \caption{The
$m-f$ phase plot of the 3D attractor.} \label{tuAtr2}
\end{figure}

Parameters $(\alpha,\beta,\gamma,\delta)$, introduced in
connection with cancer cells morphology and dynamics could also
influence the very important factor chromatin associated with
aggressive tumor phenotype and
shorter patient survival time.
\begin{figure}[ht]
\centerline{\includegraphics[width=8cm]{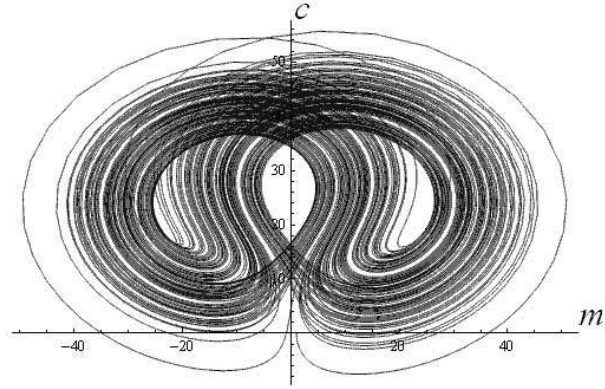}} \caption{The
$m-c$ phase plot of the 3D attractor.} \label{tuAtr3}
\end{figure}
\begin{figure}[ht]
\centerline{\includegraphics[width=6cm]{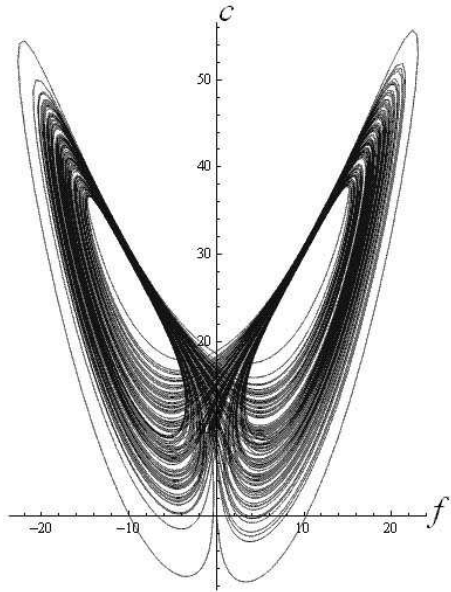}} \caption{The
$f-c$ phase plot of the 3D attractor.} \label{tuAtr4}
\end{figure}

For computations, the parameters were set to $\alpha $ = 0.06,
$\beta $ = 0.05, $\gamma $ = 26.5 and $\delta $ = 40. Small
variation of these chosen values would not affect the qualitative
behavior of the new temporal model (\ref{md1})--(\ref{md4}).
Simulations of (\ref{md1})--(\ref{md4}), using the same initial
conditions and the same non-dimensional parameters as before, show
chaotic behavior in the form of Lorenz-like strange attractor in
the 3D ($f-m-c$) subspace of the full 4D ($n-f-m-c$) phase-space
(Figures \ref{tuAtr1}--\ref{tuAtr4}).

The spatio-temporal system of rate PDEs corresponding to the
system in (15)--(18) provides the following multi-scale cancer
invasion model.
\begin{eqnarray}
\frac{\partial n}{\partial t} &=&d_{n}\nabla ^{2}n-\rho \nabla
\cdot
(n\nabla f),  \label{mpd1} \\
\frac{\partial f}{\partial t} &=&\alpha \eta (m-f),  \label{mpd2} \\
\frac{\partial m}{\partial t} &=&d_{m}\nabla ^{2}m+\kappa n-\sigma
m,
\label{mpd3} \\
\frac{\partial c}{\partial t} &=&d_{c}\nabla ^{2}c+\nu f-\omega
n-\phi c. \label{mpd4}
\end{eqnarray}

The new tumor--growth model (\ref{mpd1})--(\ref{mpd4}) retains all
the qualities of the original AC model (\ref{d1})--(\ref{d4}) plus
includes the temporal chaotic `butterfly'--attractor. This chaotic
behavior may be a more realistic view on the tumor growth, including
stochastic--like long--term unpredictability and
uncontrollability, as well as sensitive dependence of a tumor
growth on its initial conditions.

Based on the new tumor growth model (\ref{mpd1})--(\ref{mpd4}), we
have re-formulated the following generic concept of carcinogenesis
and metastasis (see scheme on the following page). With the model
of strongly coupled PDEs, remodelling of extracellular matrix
(ECM) causes a whole process of the movement of invading cells
with increased haptotaxis, and at the same time decreased enzyme
productions level. This can alter chromatin structure, which plays
an important role in initiating, propagating and terminating
cellular response to DNA damage \cite{Downs}. The effect of
haptotaxis in the process of cells invading could be modelled with
travelling-wave (Fisher) equation \cite{Marchant}.

\newpage
$\fbox{Extracellular matrix [ECM] }$

$\qquad \ \ \ \downarrow $

$\fbox{Carcinogenesis -- nonlinearly coupled diffusion PDEs}$

\qquad $\ \ \ \downarrow \qquad \qquad \qquad \qquad \qquad \qquad
\qquad \qquad \downarrow $

$\fbox{%
\begin{tabular}{c}
Mutation of p53 gene \\
``Guardian of the Genome''
\end{tabular}
}\qquad \fbox{%
\begin{tabular}{c}
Subpopulations of tumor cells initiate \\
tumor growth (cancer stem cells) \cite{Sleeman}
\end{tabular}
}$

$\qquad \ \ \ \downarrow $

{\small $\fbox{$
\begin{array}{c}
\begin{tabular}{c}
\textbf{ECM} \\
\textbf{Remodelling}
\end{tabular}
\\
\ \fbox{\
\begin{tabular}{c}
ECM \\
primary tumor cell \\
$\fbox{%
\begin{tabular}{c}
Extracellular \\
matrix
\end{tabular}
}$ \\
$\downarrow $ \\
$\fbox{%
\begin{tabular}{c}
ECM \\
mutation cell
\end{tabular}
}$%
\end{tabular}
} \\
\downarrow \text{escape} \\
\fbox{%
\begin{tabular}{c}
Spontaneous \\
cell-cell fusion
\end{tabular}
} \\
\downarrow  \\
\fbox{%
\begin{tabular}{c}
Travelling through \\
the bloodstream
\end{tabular}
} \\
\downarrow  \\
\fbox{%
\begin{tabular}{c}
Stopping at a \\
distant site
\end{tabular}
} \\
\downarrow  \\
\fbox{%
\begin{tabular}{c}
Bone \\
marrow
\end{tabular}
}
\end{array}
$}\qquad \cone{}\qquad \fbox{$
\begin{array}{c}
\begin{tabular}{c}
\textbf{Bone marrow:} \\
\textbf{Tumor--specific} \\
\textbf{premetastatic site} \\
\textbf{before metastasis} \\
(\textbf{pre-metastatic niches} \cite{Sleeman})
\end{tabular}
\\
\begin{array}{c}
\fbox{%
\begin{tabular}{c}
Bone marrow--derived \\
haematopoetic \\
progenitor cell express
\end{tabular}
} \\
\downarrow
\end{array}
\\
\fbox{%
\begin{tabular}{c}
Vascular endotelial \\
growth--factor receptor 1 \\
(VEGFR1) home \\
to tumor--specific \\
premetastatic site
\end{tabular}
} \\
\downarrow  \\
\fbox{%
\begin{tabular}{c}
Form cellar clusters \\
before the arrival \\
of tumor cells
\end{tabular}
} \\
\downarrow  \\
\fbox{%
\begin{tabular}{c}
VEGFR1 + cells \\
$\Rightarrow $ integrin $\alpha 4\beta 1$ \\
-- adhesion molecule \\
which joins the cell \\
and ECM
\end{tabular}
} \\
\mathbf{Metastasis-Chaos}
\end{array}
$}$ }

$\fbox{\textbf{Solution for cancer control:}\newline
\fbox{%
\begin{tabular}{c}
Celular retraining of cancer stem cells \cite{Janecka} \\
and/or activation of positive function of \\
cyclin-dependent kinase inhibitor p27 \cite{Chu} \\
and/or decreased expression of SATB1 \cite{Cai,Han}
\end{tabular}
}}$\\ \bigbreak

\begin{figure}[ht]
\centerline{\includegraphics[width=12cm]{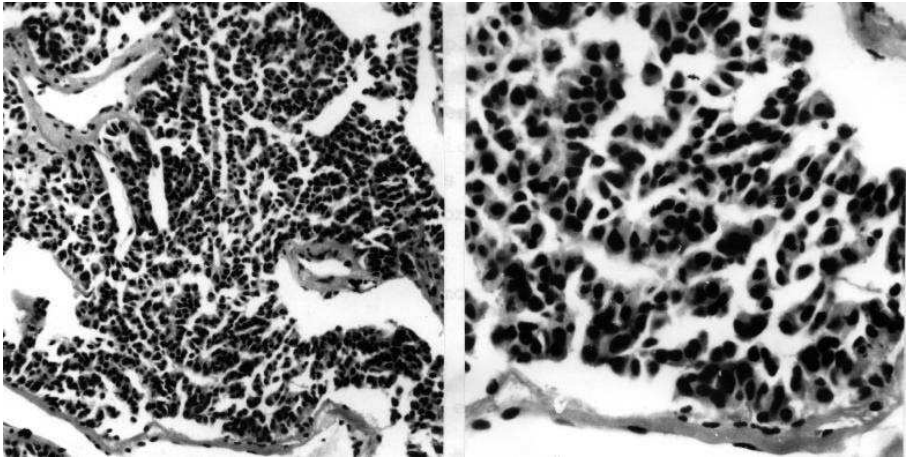}}
\caption{Carcinoma intraductale mamae: magnification at 160
$\times$ (left) and 420 $\times$ (right).} \label{Scan1}
\end{figure}

\begin{figure}[ht]
\centerline{\includegraphics[width=12cm]{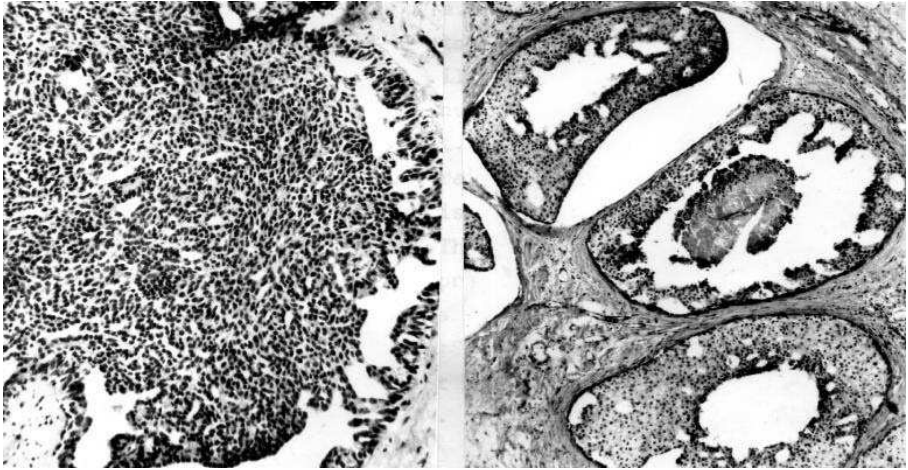}}
\caption{Carcinoma ductale in situ mamae with magnification at 80
$\times$ (left) and carcinoma comedo mamae with magnification at
420 $\times$ (right).} \label{Scan2}
\end{figure}

The proposed model (\ref{mpd1})--(\ref{mpd4}) describes chaotic
behavior relevant to the invasion of cancer cells (see Figures
\ref{Scan1} and \ref{Scan2}). As devices for controlling
metastasis/chaos we suggest the following processes:
Cellular retraining of cancer stem cells and/or activation of
positive function of cyclin-dependent kinase inhibitor p27 and/or
decreased expression of SATB1, which is correlated with aggressive
tumor phenotype in breast cancer and shorter patient survival time
\cite{Janecka,Chu,Cai,Han}.
To connect the new parameters $(\alpha,\beta,\gamma,\delta)$ to
the therapeutic regime we note the fact that significant factor of
any therapy is tumor re-growth during the rest periods between
therapy applications, which is again dependent on proliferation
fraction dynamics $\alpha $ \cite{Kozusko}.

Also, the age of the tumor may translate to different levels of
response to a standard therapy. As mentioned earlier, the number of
cells, $\gamma $, is connected with the Cdk inhibitor p27, for
which levels and activity increase in response with the cell
density $\gamma $, differentiation following loss of adhesion to
the ECM. Cdk inhibitor p27 regulates cell proliferation, cell
motility and apoptosis and is the essential element for
understanding transduction pathways in the regulation of normal
and malignant cell proliferation as well as it is new hope for
therapeutic intervention \cite{Chu}.

Introducing the parameter set $(\alpha,\beta,\gamma,\delta)$
into the model of tumor cell morphology and
function could lead to insight into the relationship between
these parameters and chromatin. By varying these parameters,
it may be possible to predict situations that result in
dangerous levels of chromatin. High levels of chromatin are
associated with cancer metastasis and some of the
most aggressive cancer types \cite{Cai,Han}.

The new temporal model (\ref{md1})--(\ref{md4}) (as well as the
corresponding spatio-temporal tumor-growth model
(\ref{mpd1})--(\ref{mpd4})) \emph{is not} sensitive to variation
of the $(\alpha,\beta,\gamma,\delta)-$values, but \emph{is}
sensitive to their corresponding places in the equations.

\section{Conclusion}

A plausible chaotic multi-scale cancer-invasion model has been
presented. The new model was formulated by introducing nonlinear
coupling into the existing hybrid multiscale Anderson-Chaplain
model. The new model describes chaotic behavior, as well as
sensitive dependence of a tumor evolution on its initial
conditions. On the basis of this chaotic tumor-growth model, a
generic concept of carcinogenesis and metastasis was described.
Effective cancer control is reflected in progressive reduction in
cancer mortality \cite{Janecka}. The proposed model suggests a
possible solution to carcinogenesis and metastasis, by combining
mathematical modelling with latest medical discoveries.\bigbreak

\noindent \textbf{Acknowledgment}\\

\noindent The authors are grateful to Professor Nada Sljapic, MD,
PhD, from the Medical School, University of Novi Sad, for expert
advice and breast cancer slides.

The first author gratefully acknowledges the support of the
Knowledge-Based Engineering Centre of University of South
Australia.

\section{Appendix}

\subsection{Vector Operators, PDEs and Diffusion Processes}

\subsubsection{The Hamilton operator $\protect\nabla$}

The \emph{core} vector differential operator $\nabla $, called
`nabla' or `del' operator, having properties analogous to those of
ordinary vectors, invented by Sir William Rowan Hamilton
(1805--1865), is defined in Cartesian ($x,y,z$)--coordinates as
\begin{equation*}
\nabla =\frac{\partial }{\partial x}\mathbf{i}+\frac{\partial }{\partial y}%
\mathbf{j}+\frac{\partial }{\partial z}\mathbf{k},
\end{equation*}
where ($\mathbf{i,j,k}$) are the unit vectors of the coordinate axes ($x,y,z$%
), respectively.

Using Hamilton's $\nabla-$operator, we can define the following
three quantities (fundamental for vector calculus and its various
applications):

\begin{description}
\item[The gradient.]  Let $\phi (x,y,z)$ define a smooth\footnote{%
`smooth' here means differentiable at each point $(x,y,z)$ in a
certain region of space} scalar field. Then the gradient of $\phi
$, denoted $\nabla \phi $ or \textsl{grad}\,$\phi $, is a
\emph{vector field} defined by
\begin{equation*}
\text{\textsl{grad}}\,\phi =\nabla \phi =\frac{\partial \phi }{\partial x}\mathbf{i}+%
\frac{\partial \phi }{\partial y}\mathbf{j}+\frac{\partial \phi }{\partial z}%
\mathbf{k}.
\end{equation*} The gradient $\text{\textsl{grad}}\,\phi$ is a vector field which points in the direction
of the greatest rate of increase of the scalar field $\phi
(x,y,z)$, and whose magnitude is the greatest rate of change.

\item[The divergence.]  Let $\mathbf{V}(x,y,z)=V_{1}\mathbf{i}+V_{2}\mathbf{j}%
+V_{3}\mathbf{k}$ define a smooth vector field. Then the divergence of $%
\mathbf{V}$, denoted $\nabla \cdot \mathbf{V}$ or \textsl{div}\,$\mathbf{V}$%
, is a \emph{scalar field} defined by
\begin{equation*}
\text{\textsl{div}}\,\mathbf{V}=\nabla \cdot \mathbf{V}=\frac{\partial V_{1}}{%
\partial x}+\frac{\partial V_{2}}{\partial y}+\frac{\partial V_{3}}{\partial
z}.
\end{equation*} The divergence $\text{\textsl{div}}\,\mathbf{V}(x,y,z)$ measures the magnitude of a vector field's \emph{source}
or \emph{sink} at a given point $(x,y,z)$. A vector field that has
zero divergence everywhere is called \emph{solenoidal.}

\item[The curl.]  If $\mathbf{V}(x,y,z)$ is a smooth vector field
as above, then the curl or rotation of $\mathbf{V}$, denoted
$\nabla \times \mathbf{V}$ or \textsl{rot}\,$\mathbf{V}$, is a
\emph{vector field} defined by the following determinant
\begin{equation*}
\text{\textsl{rot}}\,\mathbf{V}=\nabla \times \mathbf{V}=\left|
\begin{array}{ccc}
\mathbf{i} & \mathbf{j} & \mathbf{k} \\
\frac{\partial }{\partial x} & \frac{\partial }{\partial y} &
\frac{\partial
}{\partial z} \\
V_{1} & V_{2} & V_{3}
\end{array}
\right| .
\end{equation*} The curl shows a vector field's \emph{rate of rotation,} that is, the direction of the axis
of rotation and the magnitude of the rotation. It can also be
described as the \emph{circulation density.} A vector field which
has zero curl everywhere is called \emph{irrotational.}
\end{description}

\subsubsection{The Laplacian operator $\protect\nabla^2$ and the fundamental
PDEs}

The Laplacian operator is the dot--product of the Hamilton
operator by itself, that is $\nabla ^{2}=\nabla \cdot \nabla $
(or, \textsl{div\,grad}, the divergence of the gradient), defined
by
\begin{equation*}
\nabla ^{2}=\frac{\partial ^{2}}{\partial x^{2}}+\frac{\partial ^{2}}{%
\partial y^{2}}+\frac{\partial ^{2}}{\partial z^{2}}.
\end{equation*}
When applied to a smooth scalar field $\phi (x,y,z)$, the
Laplacian gives
\begin{equation}
\nabla ^{2}\phi =\text{\textsl{div\,grad}}\,\phi=\nabla \cdot (\nabla \phi )=\frac{\partial ^{2}\phi }{%
\partial x^{2}}+\frac{\partial ^{2}\phi }{\partial y^{2}}+\frac{\partial
^{2}\phi }{\partial z^{2}}.  \label{lap}
\end{equation}

If the expression (\ref{lap}) is equal to zero, we get the (elliptic) \emph{%
Laplace PDE,}\footnote{%
Note that Laplacian can also be applied to the smooth vector field $\mathbf{V%
}(x,y,z)$, producing the \emph{vector Laplace equation}, which is
important in Maxwell's electrodynamics.}
\begin{equation*}
\nabla ^{2}\phi =0\text{, \ \ \ \ \ \ \ or \ \ \ \ \ \
}\frac{\partial
^{2}\phi }{\partial x^{2}}+\frac{\partial ^{2}\phi }{\partial y^{2}}+\frac{%
\partial ^{2}\phi }{\partial z^{2}}=0.
\end{equation*}
The Laplace PDE describes various stationary fields: thermostatic,
electrostatic, magnetostatic, etc. It is always solved subject to
certain \emph{boundary conditions}.

If the expression (\ref{lap}) is proportional (with the constant
$c$) to the
first time derivative $\partial \phi /\partial t$ of the scalar field $\phi $%
, we get the (parabolic) \emph{heat PDE,}
\begin{equation}
\nabla ^{2}\phi =c\frac{\partial \phi }{\partial t}\text{, \ \ \ \
\ \ \ or \ \ \ \ \ \ }\frac{\partial ^{2}\phi }{\partial
x^{2}}+\frac{\partial
^{2}\phi }{\partial y^{2}}+\frac{\partial ^{2}\phi }{\partial z^{2}}=c\frac{%
\partial \phi }{\partial t}.  \label{heat}
\end{equation}
The heat PDE describes heat conduction, as well as diffusion
processes of various biophysical natures. It is always solved
subject to certain \emph{initial and boundary conditions}.

If the expression (\ref{lap}) is equal to the second time derivative $%
\partial ^{2}\phi /\partial t^{2}$ of the scalar field $\phi $, we get the
(hyperbolic) \emph{wave PDE,}
\begin{equation*}
\nabla ^{2}\phi =c\frac{\partial ^{2}\phi }{\partial t^{2}}\text{,
\ \ \ \ \ \ \ or \ \ \ \ \ \ }\frac{\partial ^{2}\phi }{\partial
x^{2}}+\frac{\partial
^{2}\phi }{\partial y^{2}}+\frac{\partial ^{2}\phi }{\partial z^{2}}=c\frac{%
\partial ^{2}\phi }{\partial t^{2}}.
\end{equation*}
The wave PDE describes wave processes of various biophysical
natures. It is always solved subject to certain \emph{initial and
boundary conditions}.

\subsubsection{General Diffusion PDE}

Density fluctuations in a material undergoing diffusion are
described by the \emph{diffusion PDE,}
\begin{equation}
\frac{\partial \phi }{\partial t}=\nabla \cdot \left( D(\phi ,\mathbf{r}%
)\,\nabla \phi (\mathbf{r},t)\right) ,  \label{dif}
\end{equation}
where $\phi (\mathbf{r},t)$ denotes the density of the diffusing
material at location $\mathbf{r}=(x,y,z)$ and time $t$ and $D(\phi
,\mathbf{r})$ is the collective diffusion coefficient for density
$f$ at location $\mathbf{r}$. If the diffusion coefficient $D$
depends on the density then the equation is nonlinear, otherwise
it is linear. If $D=1/c$ is constant, then the equation reduces to
the heat equation (\ref{heat}).

The diffusion equation (\ref{dif}) can be derived from the
\emph{continuity PDE,}
\begin{equation}
\frac{\partial \phi }{\partial t}+\nabla \cdot \mathbf{f}=0,\qquad \text{%
where ~}\mathbf{f}\text{$~$ is the \emph{flux} of the diffusing
material}, \label{con}
\end{equation}
which states that a change in density in any part of the system is
due to inflow and outflow of material into and out of that part of
the system. The diffusion equation (\ref{dif}) can be obtained
easily from the continuity equation (\ref{con}) when combined with
the phenomenological \emph{Fick's first law}, which assumes that
the flux of the diffusing material in any part of the system is
proportional to the local density gradient,
\begin{equation*}
\mathbf{f}=-D\,(\phi )\,\nabla \,\phi \,(\,\mathbf{r},t\,).
\end{equation*}

The general (scalar) \emph{transport PDE,}
\begin{equation}
\frac{\partial \phi }{\partial t}+\nabla \cdot
\mathbf{f}(t,\mathbf{r},\phi ,\nabla \phi
)=\mathbf{g}(t,\mathbf{r},\phi ),\qquad\text{where $\mathbf{g}$ is
called the \emph{source,}} \label{tr1}
\end{equation}
describes transport phenomena such as heat transfer, mass
transfer, fluid dynamics, etc. All the transfer processes express
a certain conservation principle. In this respect, any
differential equation addresses a certain quantity as its
dependent variable and thus expresses the balance between the
phenomena affecting the evolution of this quantity. For example,
the temperature of a fluid in a heated pipe is affected by
convection due to the solid--fluid interface, and due to the
fluid--fluid interaction. Furthermore, temperature is also
diffused inside the fluid. For a steady--state problem, with the
absence of sources, a differential equation governing the
temperature will express a balance between convection and
diffusion.

If the dependent variable (scalar or vector field) is denoted by
$\phi $, the general transport PDE (\ref{tr1}) can be rewritten as
\begin{equation}
\underbrace{\frac{\partial {\rho \phi }}{\partial t}}_{\text{Transient term}%
}+\underbrace{\nabla \cdot (\rho \mathbf{u}\phi
)}_{\text{Convection term}}=~
\underbrace{\nabla \cdot (D\nabla \phi )}_{\text{Diffusion term}}~+%
\underbrace{S_{\phi }}_{\text{Source term}}. \label{tr2}
\end{equation}
The terms in (\ref{tr2}) have the following meaning:
\begin{description}
    \item[the transient term,] $\frac{\partial{\rho \phi}}{\partial
    t},$ accounts for the accumulation of $\phi$ in the concerned control
    volume;
    \item[the convection term,] accounts for the transport of $\phi$ due to
    the existence of the velocity field (note the velocity $\mathbf{u}$ multiplying
    $\phi$);
    \item[the diffusion term,] $ \nabla \cdot (D \nabla \phi ), $ accounts for the transport of $\phi$ due to its
    gradients;
    \item[the source term,] $S_{\phi }, $ accounts for any sources or sinks that either create or destroy $\phi$.
    Any extra terms that cannot be cast into the convection or diffusion terms are considered as source terms.
\end{description}

\subsection{A paradigm of a strange attractor}

\textit{Chaos theory}, of which \textit{turbulence} is the most
extreme form, started in 1963, when Ed Lorenz from MIT took the
\textit{Navier--Stokes equations} from viscous fluid dynamics and
reduced them into three first--order coupled nonlinear ODEs, to
demonstrate the idea of sensitive dependence upon initial
conditions and associated \textit{chaotic behavior}.

\subsubsection{Turbulence and chaos theory}

Viscous fluids evolve according to the nonlinear Navier--Stokes
PDEs
\begin{equation}
\frac{\partial\mathbf{u}}{\partial t}+\mathbf{u}\cdot \nabla\mathbf{u}+{\nabla p}%
/\rho =\nu \nabla^2 \mathbf{u}+\mathbf{f},  \label{NavSt}
\end{equation}%
where $\mathbf{u}=\mathbf{u}(x^{i},t),\,(i=1,2,3)$~ is the fluid
3D velocity, $p=p(x^{i},t)$ is the pressure field, $\rho ,\nu $
are the fluid density and viscosity coefficient, while $\mathbf{f}=\mathbf{f}%
(x^{i},t)$ is the nonlinear external energy source. To simplify
the problem, we can impose to $\mathbf{f}$ the so--called
\emph{Reynolds condition}, $\langle \mathbf{f}\cdot
\mathbf{u}\rangle =\varepsilon $, where $\varepsilon $ is the
average rate of energy injection. Fluid dynamicists believe that
\emph{Navier--Stokes equations} (\ref{NavSt}) \emph{accurately
describe turbulence}.

\subsubsection{`Lorenz mask' attractor}

Turbulence is a spatio-temporal chaotic attractor. In general, an
\textit{attractor} is a set of system's states (i.e., points in
the system's phase--space), invariant under the dynamics, towards
which neighboring states in a given \textit{basin of attraction}
asymptotically
approach in the course of dynamic evolution.\footnote{%
A \emph{basin of attraction} is a set of points in the system's
phase--space, such that initial conditions chosen in this set
dynamically evolve to a particular attractor.} An attractor is
defined as the smallest unit which cannot be itself decomposed
into two or more attractors with distinct basins of attraction.
This restriction is necessary since a dynamical system may have
multiple attractors, each with its own basin of attraction (see,
e.g., \cite{StrAttr}).

Conservative systems do not have attractors, since the motion is
periodic. For dissipative dynamical systems, however, volumes
shrink exponentially, so attractors have 0 volume in $n$D
phase--space.

In particular, a stable \textit{fixed--point} surrounded by a
dissipative
region is an attractor known as a \textit{map sink}.\footnote{%
A \emph{map sink} is a stable fixed--point of a map which, in a
dissipative dynamical system, is an attractor.} Regular attractors
(corresponding to 0 \textit{Lyapunov exponents}) act as
\textit{limit cycle}\emph{s}, in which trajectories circle around
a limiting trajectory which they asymptotically
approach, but never reach. The so--called \textit{strange attractor}\emph{s}%
\footnote{%
A strange attractor is an attracting set that has zero measure in
the embedding phase--space and has fractal dimension. Trajectories
within a strange attractor appear to skip around randomly.} are
bounded regions of phase--space (corresponding to positive
Lyapunov characteristic exponents) having zero measure in the
embedding phase--space and a \textit{fractal dimension}.
Trajectories within a strange attractor appear to skip around
randomly.

In 1963, Ed Lorenz from MIT was trying to improve weather
forecasting. Using a primitive computer of those days, he
discovered the first \emph{chaotic attractor}. Lorenz used three
Cartesian variables, $(x,y,z)$, to define \textit{atmospheric
convection}. Changing in time, these variables gave him a
trajectory in a (Euclidean) 3D--space. From all starts,
trajectories settle onto a chaotic, or \textit{strange attractor}.
More precisely, Lorenz reduced the Navier--Stokes equations
(\ref{NavSt}) for \textit{convective B\'{e}nard fluid flow} into
three first order coupled nonlinear ODEs and demonstrated with
these the idea of sensitive dependence upon initial conditions and
chaos (see \cite{Lorenz,Sp}).
\begin{figure}[ht]
\centerline{\includegraphics[width=13.5cm]{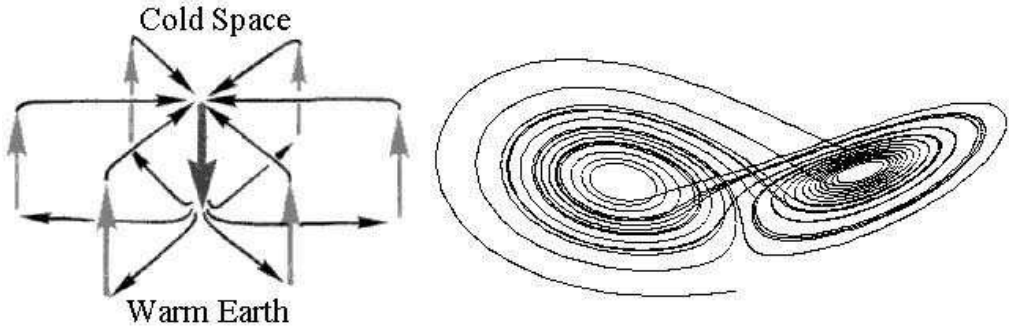}}
\caption{B\'{e}nard cells, showing a typical vortex of a rolling
air, with a warm air rising in a ring and a cool air descending in
the center (left). A simple model of a pair of B\'{e}nard cells
given by the celebrated `Lorenz--butterfly' (or, `Lorenz--mask')
\textit{strange attractor} (right) \protect\cite{ComputMind}.}
\label{Lorenz}
\end{figure}

The celebrated \emph{Lorenz ODEs} read
\begin{equation}
\dot{x}=a(y-x),\qquad \dot{y}=bx-y-xz,\qquad \dot{z}=xy-cz,
\label{LorenzSys}
\end{equation}
where $x$, $y$ and $z$ are dynamical variables, constituting the 3D \emph{%
phase--space} of the \textit{Lorenz system}; and $a$, $b$ and $c$
are the parameters of the system. Originally, Lorenz used this
model to describe the unpredictable behavior of the weather, where
$x$ is the rate of convective overturning (convection is the
process by which heat is transferred by a moving fluid), $y$ is
the horizontal temperature overturning, and $z$ is the
vertical temperature overturning; the parameters are: $a\equiv P-$%
proportional to the \textit{Prandtl number} (ratio of the fluid
viscosity of
a substance to its thermal conductivity, usually set at $10$), $b\equiv R-$%
proportional to the Rayleigh number (difference in temperature
between the top and bottom of the system, usually set at $28$),
and $c\equiv K-$a number proportional to the physical proportions
of the region under consideration
(width to height ratio of the box which holds the system, usually set at $%
8/3 $). The Lorenz system (\ref{LorenzSys}) has the properties:

\begin{enumerate}
\item  \emph{Symmetry}: $(x,y,z)\rightarrow (-x,-y,z)$ for all
values of the parameters, and

\item  The $z-$axis $(x=y=0)$ is \emph{invariant} (i.e., all
trajectories that start on it also end on it).
\end{enumerate}

Nowadays it is well--known that the Lorenz model is a paradigm for
low--dimensional chaos in dynamical systems in meteorology,
hydrodynamics, laser physics, superconductivity, electronics, oil
industry, chemical and biological kinetics, etc.

The 3D \emph{phase--portrait} of the Lorenz system (\ref{Lorenz})
shows the
celebrated `\textit{Lorenz mask}', a special type of \textit{chaotic, strange%
}, or, \textit{fractal attractor} (see Figure \ref{Lorenz}). It
depicts the famous `\textit{butterfly effect}', (i.e., sensitive
dependence on initial conditions, that is, a tiny difference in
initial conditions is amplified until two outcomes are totally
different), so that the long term behavior becomes impossible to
predict (e.g., long term weather forecasting). The Lorenz mask has
the following characteristics: (i) trajectory does not intersect
itself in three dimensions; (ii) trajectory is not periodic or
transient; (iii) general form of the shape does not depend on
initial conditions; and (iv) exact sequence of loops is very
sensitive to the initial conditions. \bigbreak


\begin{thebibliography}{9}
\bibitem{Nature} R.A. Gatenby, P.K. Maini, Mathematical oncology:
Cancer summed up. Nature, \textbf{421}, 321, (2003)

\bibitem{Roose}  T. Roose, S.J. Chapman, P.K. Maini, Mathematical Models of
Avascular Tumor Growth. SIAM Rev. \textbf{49}(2), 179--208,
(2007).

\bibitem{GaneshSprBig}  V. Ivancevic, T. Ivancevic, Geometrical Dynamics of
Complex Systems. Springer, Dordrecht, (2006)

\bibitem{GCompl}  V. Ivancevic, T. Ivancevic, Complex Dynamics: Advanced
System Dynamics in Complex Variables. Springer, Dordrecht, (2007)

\bibitem{TacaNODY} T. Ivancevic, L. Jain, J. Pattison, A. Hariz, Nonlinear Dynamics and Chaos Methods in
Neurodynamics and Complex Data Analysis. Nonl. Dyn. (Springer) (to
appear)

\bibitem{horizons} G. Guarini, E. Onofri, E. Menghetti, New horizons in medicine. The
attractors. Recenti Prog. Med. (in Italian) \textbf{84}(9),
618-623, (1993).

\bibitem{frac} R. Sedivy, Fractal tumours: their real and virtual
images. Wien Klin. Wochenschr. (in German) \textbf{108}(17),
547-551, (1996).

\bibitem{frac2} R. Sedivy, C. Windischberger, Fractal analysis of a breast carcinoma--presentation of a modern morphometric
method. Wien Klin. Wochenschr. (in German) \textbf{148}(14),
335-337, (1998).

\bibitem{Anderson98} A.R.A. Anderson, M.A.J. Chaplain, Continuous
and discrete mathematical models of tumor-induced angiogenesis.
Bul. Math. Biol. \textbf{60}, 857-000, (1998).

\bibitem{Anderson00} A.R.A. Anderson, M.A.J. Chaplain, E.L. Newman, R.J.C.
Steele, A.M. Thompson, Mathematical modelling of tumour invasion
and metastasis. J. Theor. Med. \textbf{2}, 129–154, (2000).

\bibitem{Anderson05} A.R.A. Anderson, A hybrid mathematical model of solid
tumour invasion: The importance of cell adhesion. Math. Med. Biol.
\textbf{22}, 163--186, (2005).

\bibitem{Anderson-Cell} A.R.A. Anderson, A.M. Weaver, P.T.Cummings, V.
Quaranta, Tumor Morphology and Phenotypic Evolution Driven by
Selective Pressure from the Microenvironment. Cell \textbf{127},
905--915, (2006).

\bibitem{Anderson06} M.A.J. Chaplain, S.R. Mc Dougal, A.R.A.
Anderson, Mathematical modelling of tumor-induced angiogenesis.
Ann. Rev. Biomed. Eng. \textbf{8}, 233-257, (2006).

\bibitem{AndersonMBE} H. Enderling, A.R.A. Anderson, M.A.J. Chaplain,
Visualisation of the numerical solution of partial differential
equation systems in three space dimensions and its importance for
mathematical models in biology. Math. Biosci. Eng. \textbf{3}(4),
571-582, (2006).

\bibitem{AndersonJTB} H. Enderling, M.A.J. Chaplain, A.R.A. Anderson, J.S.
Vaidya, A mathematical model of breast cancer development, local
treatment and recurrence. J. Th. Biol. \textbf{246}, 245–259,
(2007).

\bibitem{Anderson08} I. Ramis-Conde, M.A.J. Chaplain, A.R.A.
Anderson, Mathematical modelling of cancer cell invasion of
tissue. Math. Comp. Mod. \textbf{47}, 533-545, (2008).

\bibitem{Chaplain08} A. Gerish, M.A.J. Chaplain, Mathematical modelling of cancer cell invasion of
tissue: Local and non-local models and the effect of adhesion. J.
Th. Biol. \textbf{250}, 684-704, (2008).

\bibitem{Marchant} B.P. Marchant, J. Norbury, J.A. Sherratt, Travelling wave solutions to a haptotaxis-dominated
model of malignant invasion. Nonlinearity, \textbf{14}, 1653–1671,
(2001)

\bibitem{OGY} E. Ott, C. Grebogi, J.A. Yorke, Controlling chaos.
Phys. Rev. Lett., \textbf{64}, 1196--1199, (1990).

\bibitem{StrAttr}  V. Ivancevic, T. Ivancevic, High--Dimensional Chaotic and
Attractor Systems. Springer, Berlin, (2006).

\bibitem{Complexity}  V. Ivancevic, T. Ivancevic, Complex
Nonlinearity: Chaos, Phase Transitions, Topology Change and Path
Integrals, Springer, Series: Understanding Complex Systems,
Berlin, (in press).

\bibitem{Lorenz}  E.N. Lorenz, Deterministic Nonperiodic Flow. J. Atmos.
Sci., \textbf{20}, 130--141, (1963).

\bibitem{Sp}  C. Sparrow, The Lorenz Equations: Bifurcations, Chaos and
Strange Attractors. Springer, New York, (1982).

\bibitem{NeuFuz}  V. Ivancevic, T. Ivancevic, Neuro-Fuzzy Associative Machinery
for Comprehensive Brain and Cognition Modelling. Springer, Berlin,
(2007)

\bibitem{ComputMind}  V. Ivancevic, T. Ivancevic, Computational Mind: A
Complex Dynamics Perspective. Springer, Berlin, (2007).

\bibitem{Sleeman} J.P. Sleeman, N. Cremers, New concepts in breast cancer metastasis:
tumor initiating cells and the microenvironment. Clin. Exp.
Metastasis. \textbf{24}(8), 707--15, (2007).

\bibitem{Janecka} I.P. Janecka, Cancer control through principles of systems science, complexity, and
chaos theory: A model. Int. J. Med. Sci. \textbf{4}(3), 164-173,
(2007).

\bibitem{Downs} J.A. Downs, M.C. Nussenzweig, A. Nussenzweig, Chromatin dynamics and the preservation
of genetic information. Nature, \textbf{447}, 951-958, 21 June
2007.

\bibitem{Chu} I.M. Chu, L. Hengst, J.M. Slingerland, The Cdk inhibitor p27 in human
cancer: prognostic potential and relevance to anticancer therapy.
Nature Rev. Cancer, \textbf{8}, 253-287, April 2008.

\bibitem{Cai} S. Cai, C.C. Lee, T. Kohwi-Shigematsu, SATB1 packages
densely looped, transcriptionally active chromatin for coordinated
expression of cytokine genes. Nature Genet. \textbf{38},
1278-1288, (2006).

\bibitem{Han} H.-J. Han, J. Russo, Y. Kohwi, T. Kohwi-Shigematsu, SATB1 reprograms gene
expression to promote breast tumor growth and metastasis. Nature,
\textbf{452}, 187–193, (2008).

\bibitem{Olive} Olive, P.L., Durand, R.E., Drug and radiation resistance in spheroids: cell
contact and kinetics. \textit{Cancer Metastasis} \textit{Rev}.
\textbf{13}, 121, (1994).

\bibitem{Kozusko} Kozusko, F., Bourdeau, M., A unified model of sigmoid tumour
growth based on cell proliferation and quiescence, Cell Prolif.
\textbf{40}(6), 824-834, (2007).

\bibitem{Castro} Castro, M.A.A., Klamt, F., Grieneisen, V.A.,
Grivicich, I., Moreira, J.C.F., Gompertzian growth pattern
correlated with phenotypic organization of colon carcinoma,
malignant glioma and non-small cell lung carcinoma cell lines.
Cell Prolif. \textbf{36}(2), 65-73, (2003).

\bibitem{Ingber} Ingber, D.E., Fibronectin controls capillary endothelial cell growth by
modulating cell shape. Proc. Natl. Acad. Sci. USA \textbf{87},
3579, (1990).

\bibitem{Boudreau} Boudreau, N., Jones, P.L., Extracellular matrix and integrin signaling: the
shape of things to come. Biochem. J. \textbf{339}, 481, (1999).
\end{thebibliography}
\end{document}